\documentclass[aps,prd,onecolumn,amsmath,amssymb]{revtex4-2}

\usepackage{amsmath}
\usepackage{amssymb}
\usepackage{amsthm}
\usepackage{graphicx}
\usepackage{hyperref}
\usepackage{xcolor}
\usepackage{booktabs}
\usepackage{aas_macros}
\usepackage[final]{changes}

\begin{document}

\title{Spectroscopy as a Probe of Spacetime Foam}

\author{A.\ G.\ Kim}
\email{agkim@lbl.gov}
\affiliation{Physics Division, Lawrence Berkeley National Laboratory, Berkeley, California 94720, USA}
\author{P.\ E.\ Nugent}
\affiliation{Applied Mathematics and Computational Research Division, Lawrence Berkeley National Laboratory, Berkeley, California 94720, USA}
\author{L.\ Wang}
\affiliation{Department of Physics and Astronomy, 4242 TAMU, College Station, Texas 77843-4242, USA}

\date{\today}

\begin{abstract}
Quantum-gravitational spacetime foam varying in space and in time
imprints, respectively, a transverse phase decoherence and an achromatic,
stochastic arrival-time jitter on light from distant
sources. We show that
the jitter
modifies the first-order temporal coherence $g^{(1)}(\tau)$ and hence the
observed spectrum, providing a spectroscopic probe of the temporal
variation, complementary to existing
imaging-channel constraints on the spatial variation. In the
dynamical-foam limit, where the foam decorrelates within the light's
coherence time,
the spectrum splits into a coherent line core suppressed
by $e^{-\sigma_\phi^2}$, with $\sigma_\phi^2$ the variance of the foam-induced
phase jitter, and an incoherent pedestal, yielding a spectral
shape-distortion bound on the foam exponent $\alpha$.
Three probes follow: the outright survival of broadband flux, the
fractional width of a spectral line, and a chromatic deficit across a
broadband continuum.
Applied to the MeV emission lines of GRB\,221009A we obtain
$\alpha \gtrsim 0.63$, a weak, lever-arm-dominated floor.
For the broadband CMB the foam induces an $e^{-\sigma_\phi^2(\nu)}$ deficit
($\sigma_\phi^2 \propto \nu^2$); a fit to
the COBE/FIRAS monopole spectrum reaches $\alpha \gtrsim 0.52$
($95\%$ CL), and next-generation spectrometers reach only
$\alpha\gtrsim 0.54$--$0.55$ (forecast), limited by the foam--temperature
degeneracy.  The pedestal-independent line-width and CMB bounds are floors below
the holographic value $\alpha = 2/3$, consistent with and excluding no part of
the imaging window reported for GRB\,221009A; only the conditional TeV
survival bound ($\alpha\gtrsim0.74$), which holds only for
near-Planck-refresh foam ($\tau_{\rm foam}\lesssim4\times10^{-28}$\,s), would
exclude it --- and, with it, the window.
All bounds assume foam that decorrelates within the light's coherence
time; they vanish in the quasi-static or long-temporal-correlation limits.
\end{abstract}

\maketitle

\section{Introduction}
\label{sec:intro}

A long-standing expectation from the union of quantum mechanics and general
relativity is that the smooth pseudo-Riemannian manifold of classical spacetime
breaks down at the Planck length
$\ell_{\rm P} = \sqrt{\hbar G/c^3} \simeq 1.616\times 10^{-35}$~m.
In Wheeler's original ``spacetime foam'' picture~\cite{Wheeler1955}, quantum
fluctuations of the metric on this scale render the geometry stochastic;
similar conclusions are reached in the Euclidean path-integral
treatment~\cite{Hawking1978} and in modern holographic and entropic-gravity
arguments from very different starting points.
While a complete theory of quantum gravity remains elusive, the question of
whether residual imprints of Planck-scale geometric fluctuations can accumulate
over cosmological baselines and become observable is one of the few empirical
handles on physics in the Planck regime, and it has motivated a substantial
body of phenomenological work over the past two
decades~\cite{Carlip2023,NgPerlman2022}.

A widely studied parametrization due to Ng and
collaborators~\cite{Ng2001,Ng2003,NgPerlman2022}, the $\alpha$-foam model,
encodes the foam dynamics in a power-law path-length variance
\begin{equation}
  \sigma_\ell^2(D) = A_\alpha\,D^{2(1-\alpha)}\,\ell_{\rm P}^{2\alpha},
  \label{eq:NgPerlmanVar}
\end{equation}
where $D$ is the line-of-sight comoving distance to the source --- the
appropriate measure rather than the luminosity distance, since the phase
accumulation is a property of spacetime itself rather than the photon
path~\cite{NgPerlman2022} --- $\alpha$ parametrizes the foam model, and
$A_\alpha$ is a dimensionless coefficient of order unity.
Ng and Perlman give the foam fluctuation equivalently in time-interval form,
$\delta t \gtrsim t^{1-\alpha}\,t_{\rm P}^{\alpha}$ with Planck time
$t_{\rm P}=\ell_{\rm P}/c$~\cite{NgPerlman2022}; since the observable in this work
is temporal, that is the more directly relevant statement, recovered explicitly
in Eq.~\eqref{eq:jitter} below.
A photon traversing distance $D$ passes through approximately
$N \sim D/\ell_{\rm P}$ independent Planck-scale spacetime cells, each imparting
a tiny random displacement to the photon path; the exponent $\alpha$ controls
how the per-cell displacements accumulate: for random-walk foam ($\alpha=1/2$),
contributions add in quadrature ($\sigma_\ell \propto \sqrt{D\ell_{\rm P}}$);
holographic foam ($\alpha=2/3$) grows more slowly, motivated by the heuristic
that the information in a spatial region scales with its bounding area rather
than its volume --- an analogy to the holographic entropy bound;
and
the weak-foam limit ($\alpha=1$) yields a distance-independent
$\sigma_\ell \propto \ell_{\rm P}$, corresponding to perfectly correlated cell
contributions.

The primary observational probe of stochastic foam to date has been the imaging
channel. Light propagating through spacetime foam accumulates a phase dispersion
$\Delta\phi \simeq 2\pi D^{1-\alpha}\ell_{\rm P}^{\alpha}/\lambda$ across the
entrance aperture, degrading PSF sharpness and erasing interference fringes once
$\Delta\phi \gtrsim 2\pi$~\cite{LieuHillman2003}.
For PKS\,1413$+$135 ($D = 1.216$~Gpc) observed with HST at $\lambda=1.6\,\mu$m,
the random-walk model ($\alpha=1/2$) predicts $\Delta\phi \sim 10\times 2\pi$,
so the detection of Airy rings in this image rules out that model;
the holographic model ($\alpha=2/3$) predicts $\Delta\phi \sim 10^{-9}\times 2\pi$
and is obviously not ruled out~\cite{NgPerlman2022,Christiansen2011,Perlman2015}.
Models with long-range transverse correlations evade even the random-walk
bound~\cite{LeeZurekChen2024}.

The most luminous GRB ever recorded, GRB\,221009A at $z=0.151$~\cite{Ravasio2024}, has been
examined in the imaging channel by Steinbring~\cite{Steinbring2023} and
Steinbring \& Ng~\cite{SteinbringNg2026}, both building on that
author's earlier treatment of high-energy instrumental response, in which
telescopes far from their diffraction limits retain only photons within the
instrumental acceptance~\cite{Steinbring2007,Steinbring2015}.
Using a multiwavelength average of foam-induced angular broadening ---
analogous to a seeing disk from atmospheric turbulence ---
Steinbring~\cite{Steinbring2023} showed that the holographic ($\alpha=2/3$)
model predicts an angular widening below $\sim\!1^{\circ}$, consistent
with the Fermi and Swift localization data for GRBs at $z\leq 5$ up to
peak energies of $\sim\!10\,\mathrm{MeV}$.
Steinbring \& Ng~\cite{SteinbringNg2026} extended this to the full
multiwavelength follow-up of GRB\,221009A --- from optical and near-infrared
through X-rays and the highest-energy gamma-rays ever attributed to a GRB ---
computing the foam-widened PSF for real
telescopes.

References~\cite{Steinbring2023,SteinbringNg2026} \replaced{advance a
stronger interpretation of those data}{read those data more strongly}.
With $\alpha\leq0.65$ already excluded by images of distant galaxies and
AGNs, the foam-widened PSF is bracketed between two instrumental scalings ---
$\alpha=0.650$, scaled to the field of view, and $\alpha=0.735$, the
weakest-blurring case, scaled to the nominal resolution --- with the
holographic $\alpha=2/3$ inside the bracket.  Across that bracket those authors
report the LAT images to be not diffraction-limited but to scale as
foam-induced blurring predicts, and \replaced{interpret}{read} the resulting
halo as attributable to foam.  The mechanism is chromatic: the delay is achromatic but the phase
it produces is not, so a shorter wavelength turns the same path-length jitter
into a larger phase and the bluest photons are blurred
most~\cite{Steinbring2007}.  Accounting for their loss beyond the
instrumental acceptance is what permits a larger $\alpha$ at fixed image
sharpness.

We develop a framework based on the first-order temporal coherence
function $g^{(1)}(\tau)$ of light from astrophysical sources.
This function is complementary to the first-order spatial coherence
function $g^{(1)}(B)$, which forms the basis of imaging-channel probes.
The first-order coherence --- the field--field autocorrelation ---
inherits a foam suppression factor
that decreases continuously from unity toward complete fringe erasure as
the accumulated phase variance $\sigma_\phi^2$ grows.
In our framework the phase variance arises from temporal fluctuations of
foam, whereas in the imaging channel it arises from spatial
fluctuations; correspondingly, foam imprints itself on the
spectrum rather than on the spatial brightness profile.

In our analysis, the foam acts on the electric field: foam modulates the
first-order temporal coherence $g^{(1)}(\tau)$ along a fixed line of sight,
and the observable effect is a spectral shape distortion.  Foam varying in
both space and time will scatter light in angle as well as distort it in
frequency; the angular channel is the imaging literature reviewed above.
This article focuses on spectral distortion and does not model deflection out
of the instrumental field of view.  It need not, because deflection drops out
of the spectral observables used here.  For the GRB line the observable is
the fractional width, a ratio in which flux lost from the beam cancels, and
what a core-only instrument records is the coherent core itself
(Appendix~\ref{app:extended}).
For the CMB, angular scattering cannot change the monopole: an isotropic,
flux-conserving redistribution convolves the sky with a normalized kernel
whose harmonic window has $W_0=1$ identically, so the $\ell=0$ amplitude is
invariant frequency by frequency; and what scatters into a sightline has been
displaced in frequency by $\sim1/\tau_{\rm foam}$, hence out of band.

The spectroscopic channel has been considered previously.
Cooke \textit{et al.}~\cite{Cooke2020} postulate a convolution kernel,
a heuristic that the authors note has not been demonstrated by a model of
spacetime quantization.
Their bound comes from a narrow
Fe\,\textsc{ii} absorption line from a quiescent cloud at $z\simeq2.34$
with VLT/ESPRESSO which ruled out $\alpha\leq0.634$.
That bound nearly coincides with the line-width floor obtained below,
$\alpha\gtrsim0.63$, despite seven decades in photon energy --- a reflection
of the lever-arm domination of such bounds: here the energy gain enters only
as $+15$ in $\ln(2\pi D_C/\lambda_{\rm obs})$ against
$\ln(D_C/\ell_{\rm P})\approx140$, and is largely offset by the fractional
width's second-order response to the jitter
(Section~\ref{sec:grb_width}).
What distinguishes our work is the physical assumptions and the details of the predictions.
We derive the distortion from the
first-order coherence of the electric field propagating through a fluctuating metric
--- the light-cone fluctuations of
Refs.~\cite{Ford1995,FordSvaiter1996}.
The two models differ in their predicted effect on spectra.
The Cooke \textit{et al.} kernel is a Gaussian of width comparable to the
line, so the displaced flux stays in the line.  In this article, flux is
redistributed into a core $e^{-\sigma_\phi^2}S_{\rm src}(\nu)$ retaining the
source profile --- Gaussian for the line considered here --- at reduced
weight, and a pedestal component of width $\sim1/\tau_{\rm foam}$ carrying
the complementary $1-e^{-\sigma_\phi^2}$.  An important element of our
observable is a flux deficit, $1-e^{-\sigma_\phi^2(\nu)}$ with
$\sigma_\phi^2\propto\nu^2$.  Within the core the absolute width in fact
narrows, so broadening survives only in the fractional width and only at
second order, $r_{\rm foam}\propto r_{\rm int}^2\,\sigma_\phi$.
Our derivation hinges on the delay varying within the light's coherence
time; in the quasi-static limit the effect vanishes entirely, a regime on
which the model of Ref.~\cite{Cooke2020}, having no time dependence, is
silent.

Because the foam variance enters the resulting bounds only through the
doubly-logarithmic factor $\ln(D_C/\ell_P)\approx138$--$141$ (GRB to CMB),
the bounds are set primarily by source geometry and probe energy; the data enter as a small
logarithmic correction (a decade of measurement precision shifts $\alpha$
by $\lesssim0.02$), so they are robust but cannot be appreciably tightened
by better data of the same kind.

The remainder of the paper is organized as follows.
Section~\ref{sec:model} introduces the spacetime foam model: the
arrival-time jitter parameterization, the temporal autocorrelation
$C_\alpha(\tau)$, and the distribution of differential jitter.
Section~\ref{sec:g1} derives the first-order coherence function
$g^{(1)}_{\rm obs}(\tau)$ of the observed field and establishes the
Gaussian closed form.
Section~\ref{sec:spec} translates these results into spectroscopic observables
via the Wiener--Khinchin theorem, establishing the split of the observed
spectrum into a coherent line core suppressed by $e^{-\sigma_\phi^2}$ and a
broad pedestal, and connects them to the imaging Strehl ratio.
Section~\ref{sec:sources} applies these three probes to two
sources: GRB\,221009A (Section~\ref{sec:grb221009a}), through the survival of
its TeV photons and the width of its MeV line, and the CMB as observed by
COBE/FIRAS (Section~\ref{sec:CMB}).
Section~\ref{sec:conclusions} summarizes the main results, sets them
against the imaging-channel literature, and outlines future
observational directions.
Table~\ref{tab:notation} collects the principal symbols used throughout.

% ==============================================================
\section{Spacetime Foam Model}
\label{sec:model}
% ==============================================================

\subsection{Arrival-Time Jitter}
\label{sec:metric_origin}

We adopt the $\alpha$-foam picture of spacetime foam: stochastic,
achromatic path-length fluctuations whose cumulative effect along a line
of sight is an arrival-time jitter $\delta t(t)$, which
\replaced{induces a phase jitter $\sigma_\phi$ in a carrier}{a carrier reads
as a phase jitter $\sigma_\phi$}.
Following the prior foam literature we take a point source on a single
line of sight.

We model the detected field as a stochastically time-warped version of the
intrinsic source field:
\begin{equation}
  E_{\rm obs}(t) = E_{\rm src}\!\left(t - \delta t(t)\right),
  \label{eq:Eobs}
\end{equation}
where $E_{\rm src}(t)$ is a single mode of a stationary, zero-mean complex
Gaussian (chaotic) field --- appropriate for the thermal and incoherent
sources here (GRB high-energy emission, CMB) --- and $\delta t(t)$ is the
foam jitter, a stationary process independent of $E_{\rm src}$.
Equation~\eqref{eq:Eobs} warps only the time argument of the field on
a fixed line of sight, and the ensemble-mean intensity is preserved,
$\langle|E_{\rm obs}|^2\rangle=\langle|E_{\rm src}|^2\rangle$.
Here $t$ is the observer's time, with the deterministic
light-travel delay and cosmological time dilation absorbed into the argument
of $E_{\rm src}$, so $\delta t(t)$ is the zero-mean foam fluctuation about
the mean arrival time.  The jitter arises from cumulative path-length
fluctuations along the line of sight of propagation distance $D$:
\begin{equation}
  \delta t(t) = \frac{1}{c}\,\Delta\ell_{\rm tot}(t).
  \label{eq:jitter}
\end{equation}
Equivalently, the rms jitter
$\sigma_{\delta t}=\sigma_\ell(D)/c=\sqrt{A_\alpha}\,t^{1-\alpha}t_{\rm P}^{\alpha}$,
with light-travel time $t=D/c$ and Planck time $t_{\rm P}=\ell_{\rm P}/c$, is
exactly the time-interval fluctuation of Ng and Perlman~\cite{NgPerlman2022} ---
the temporal form on which the present analysis rests.

This form is motivated by results on light-cone fluctuations.
Ford~\cite{Ford1995} showed that
metric fluctuations from a graviton bath smear the light-cone singularity of
the field's two-point functions, the metric-averaged retarded Green's
function acquiring Gaussian support both inside and outside the classical
light cone --- some photons arriving early, others late.  Ford and
Svaiter~\cite{FordSvaiter1996} carried this to the two-time level, showing
that flight times are Gaussian-distributed about the classical value and
that photons emitted in rapid succession have correlated flight
times, for which they derive a correlation function.
Equation~\eqref{eq:Eobs}, with a stationary Gaussian $\delta t(t)$ of
correlation $C_\alpha(\tau)$, is the field-level encapsulation of those two
results.

Because the path-length fluctuation is geometric---arising from stochastic
metric perturbations of the FLRW background rather than a dispersive
medium---$\delta t$ is achromatic (frequency-independent), distinguishing foam
from plasma dispersion, interstellar scintillation, and standard spectral
distortions~\cite{NgPerlman2022}.  \replaced{In a monochromatic carrier of
observed frequency $\omega_0=2\pi\nu_0$ the same delay induces}{A
monochromatic carrier of observed frequency $\omega_0=2\pi\nu_0$ reads the
same delay as} a phase jitter
$\delta\phi=\omega_0\,\delta t$, giving a phase variance
\begin{equation}
  \sigma_\phi^2 \;=\; \left(\frac{2\pi}{\lambda_{\rm obs}}\right)^{\!2}
  \sigma_\ell^2(D_C).
  \label{eq:sigmaphi_cosmo}
\end{equation}

This is the phase accumulation along a single ray, or in a transverse patch in which the foam
is coherent.
\citet{ChenWenMa2014} point out that the phase dispersion is damped relative
to the path-length dispersion for light propagating through a volume of
incoherent foam.  The field samples a three-dimensional region rather than
following a ray, and transverse structure finer than $\lambda$ scatters only
into evanescent modes, so that for cells uncorrelated beyond $\ell_{\rm P}$
the accumulated phase amplitude carries an extra factor
$\ell_{\rm P}/\lambda$ --- some $5\times10^{-22}$ at $37$\,MeV, and its
square in the variance.  The effect thus depends on the transverse
correlation length of the foam, which
Eq.~\eqref{eq:NgPerlmanVar} does not fix and which we do not model here.
\citet{LeeZurekChen2024} work through a specific case, the pixellon
realization of the Verlinde--Zurek effect, in which holographic degrees of
freedom on the boundary of a causal diamond drive fluctuations in its size;
the accumulated variance matches the random walk, $\langle\Delta L^2\rangle
\sim\ell_{\rm P}D$, but the power sits in the low angular harmonics rather
than at $1/\ell_{\rm P}$.  They find the suppression absent for such
fluctuations, whose power lies at wavenumbers the field can carry.  We presume
Eq.~\eqref{eq:sigmaphi_cosmo}, as is often done in the $\alpha$-foam literature,
but it is important to remember
that transverse correlation structure can affect a purely temporal
observable: the filtering acts on $\sigma_\phi^2$ itself rather than
on its transverse gradient, and so reaches the spectrum as directly as it
reaches the image.

\subsection{Temporal Decorrelation of Spacetime Foam}
\label{sec:Calpha_origin}

The foam variance $\sigma_\ell^2(D)$ (Eq.~\eqref{eq:NgPerlmanVar})
fixes the amplitude of the path-length fluctuations along a single
ray but not their temporal correlation.  Quantum-foam models posit
that each Planck-scale cell fluctuates on a characteristic timescale
$\tau_{\rm foam}$, so the jitter at times $t$ and $t+\tau$ is correlated
for $\tau\ll\tau_{\rm foam}$ and essentially independent for
$\tau\gg\tau_{\rm foam}$.  We encode this in the normalized temporal
autocorrelation $C_\alpha(\tau;D)$ of the integrated path-length
fluctuation $\Delta\ell_{\rm tot}(t)$,
\begin{equation}
  \big\langle \Delta\ell_{\rm tot}(t)\,
              \Delta\ell_{\rm tot}(t+\tau) \big\rangle
  = \sigma_\ell^2(D)\,C_\alpha(\tau;D),
  \label{eq:pathcov1ray}
\end{equation}
normalized so that $C_\alpha(0;D)=1$.  Because $C_\alpha$ is defined on
the integrated path, its temporal shape inherits a residual
$D$-dependence from the way the per-cell foam dynamics over the
$\sim D/\ell_{\rm P}$ cells crossed by the line of sight combine; the
explicit $D$-argument records this dependence.

The underlying foam correlation is spatio-temporal,
$C(\mathbf{r},\tau)$, and $C_\alpha(\tau;D)$ is its effective temporal part
along the ray. The angular extent of the sources does not enter, since for an incoherent
source the sightlines superpose in intensity and the foam is never averaged
across them (Appendix~\ref{app:extended}).\   The reduction to a
purely temporal $C_\alpha$ assumes the spatio-temporal correlation
factorizes, $C(\mathbf{r},\tau)\approx C_{\rm sp}(\mathbf{r})\,C_\alpha(\tau)$,
with the spatial factor absorbed into the normalization; we flag this as an
assumption, since the spatial correlation structure over cosmological path
lengths is not constrained here.  Non-factorization would change how the per-cell
dynamics combine along the path, and hence the shape of $C_\alpha$ at finite
lag; its long-lag limit, $C_\alpha\to0$, would be untouched.  That
distinction matters: of the observables developed in
Section~\ref{sec:spec}, some depend on the finite-lag shape and some only on
the long-lag limit.\ 

The timescale $\tau_{\rm foam}$ is an additional free parameter of the
foam model, not fixed by the $\alpha$-model variance relation alone.
In the standard phenomenological framework~\cite{AmelinoCamelia2013,Ng2003},
the foam refreshes on the Planck time
$\tau_P \equiv \ell_P/c \approx 5\times10^{-44}$~s, so the integrated
path-length jitter $\Delta\ell_{\rm tot}(t)$ --- a sum of
$\sim D/\ell_P$ independent contributions --- likewise decorrelates on
timescale $\tau_{\rm foam}\sim\tau_P$.  This is the dynamical-foam
regime adopted throughout: $\tau_{\rm foam}$ is far shorter than any
timescale on which any observable coherence has support, so
$C_\alpha(\tau)\to 0$ there.  By contrast, models with long-range
temporal correlations --- the temporal analogue of the long-range
transverse spatial correlations of Ref.~\cite{LeeZurekChen2024}
that evade interferometric imaging bounds --- would give
$C_\alpha(\tau)\approx 1$ over macroscopic lags and drive
$\sigma_\Delta^2(\tau)\to 0$, suppressing the foam signal; the bounds
derived below therefore rest on the dynamical-foam assumption.

\subsection{Differential Jitter and Its Distribution}
\label{sec:differential_jitter}

We now define the differential jitter, which enters the coherence
integrals, between photons arriving at times $t$ and $t+\tau$:
\begin{equation}
  \Delta(\tau)
  \equiv \delta t(t+\tau) - \delta t(t).
  \label{eq:DeltaExt}
\end{equation}
Expanding the square and applying Eq.~\eqref{eq:pathcov1ray}:
\begin{align}
  \sigma_\Delta^2(\tau)
  &= \frac{1}{c^2}\big\langle
    [\Delta\ell_{\rm tot}(t+\tau)
     - \Delta\ell_{\rm tot}(t)]^2
    \big\rangle \notag \\
  &= \frac{1}{c^2}\left[
    \sigma_\ell^2 + \sigma_\ell^2
    - 2\sigma_\ell^2\,C_\alpha(\tau)
    \right] \notag \\[4pt]
  &= \frac{2\sigma_\ell^2(D)}{c^2}
    \left[1 - C_\alpha(\tau;D)\right].
  \label{eq:diffVarExt}
\end{align}

By the central limit theorem (CLT), $\Delta(\tau)$
is a sum of contributions from many foam cells and is therefore Gaussian with zero mean:
\begin{equation}
  \Delta(\tau)
  \sim \mathcal{N}\!\left(0,\,\sigma_\Delta^2(\tau)\right),
  \label{eq:kernel}
\end{equation}
where $\sigma_\Delta^2$ is taken from Eq.~\eqref{eq:diffVarExt}; for
any source at $D \gtrsim 1$~kpc, the cell count
$N_{\rm cells} \sim D/\ell_{\rm P} \gtrsim 10^{54}$ amply satisfies
the large-$N$ requirement and higher-cumulant corrections are negligible.  For $\alpha = 1/2$ (random walk), per-cell path-length variances
are equal ($\ell_{\rm P}^2$ each) and the CLT applies in its standard form.
For $\alpha \neq 1/2$, the per-cell contributions scale differently with $D$
and the equality of per-cell variances is a model assumption rather than a
direct CLT consequence; and at the correlated ($\alpha\to1$) end the
per-cell increments are not independent, so the large-$N$ argument does not
apply. Gaussianity is then adopted as a modeling choice, minimal and
consistent with the $\alpha$-model.

% ==============================================================
\section{First-Order Coherence}
\label{sec:g1}
% ==============================================================

\subsection{Jitter Convolution Formula for $g^{(1)}$}

The observed first-order coherence is
\begin{equation}
  g^{(1)}_{\rm obs}(\tau) = \frac{\langle E^*_{\rm obs}(t)\,E_{\rm obs}(t+\tau)\rangle}
                                  {\langle|E_{\rm obs}(t)|^2\rangle}.
  \label{eq:g1def}
\end{equation}
Substituting Eq.~\eqref{eq:Eobs} and letting
$u = t - \delta t(t)$, $v = t+\tau - \delta t(t+\tau)$, so that
$v - u = \tau - \Delta(\tau)$, stationarity of $E_{\rm src}$ gives
\begin{equation}
  \langle E^*_{\rm src}(u)\,E_{\rm src}(v)\rangle
  = \langle|E_{\rm src}|^2\rangle\,g^{(1)}_{\rm src}(\tau - \Delta(\tau)).
\end{equation}
Since $\delta t$ is independent of $E_{\rm src}$ and
$\langle|E_{\rm obs}|^2\rangle=\langle|E_{\rm src}|^2\rangle$ by stationarity
(so the Eq.~\eqref{eq:g1def} normalization cancels), averaging over the
jitter yields
\begin{equation}
  g^{(1)}_{\rm obs}(\tau)
    = \big\langle g^{(1)}_{\rm src}(\tau - \Delta(\tau))\big\rangle_{\delta t}
    = \int_{-\infty}^{\infty} \mathcal{N}(\Delta;0,\sigma_\Delta^2(\tau))\,
      g^{(1)}_{\rm src}(\tau - \Delta)\,d\Delta.
  \label{eq:g1obs}
\end{equation}
For a quasi-monochromatic source $g^{(1)}_{\rm src}(\tau)=e^{-i\omega_0\tau}\,
\gamma(\tau)$ with $\gamma$ the slowly varying envelope coherence,
Eq.~\eqref{eq:g1obs} reads
\begin{equation}
  g^{(1)}_{\rm obs}(\tau)
    = e^{-i\omega_0\tau}\,
      \big\langle e^{\,i\omega_0\Delta(\tau)}\,
                  \gamma(\tau-\Delta(\tau))\big\rangle_{\delta t},
  \label{eq:g1obs_carrier}
\end{equation}
the envelope responding to the achromatic differential delay $\Delta(\tau)$
(variance $\sigma_\Delta^2$) and the carrier acquiring the phase
$\omega_0\Delta(\tau)$.  The carrier suppression is the characteristic function
of that phase,
\begin{equation}
  \mathcal{F}_{\rm foam}(\tau)\equiv\big\langle e^{\,i\omega_0\Delta(\tau)}\big\rangle_{\delta t}
  = e^{-\omega_0^2\sigma_\Delta^2(\tau)/2}
  = e^{-\sigma_\phi^2\,[1-C_\alpha(\tau;D)]},
  \label{eq:Ffoam}
\end{equation}
with $\mathcal{F}_{\rm foam}(0)=1$, the differential phase variance
$D_\phi(\tau)\equiv\omega_0^2\sigma_\Delta^2(\tau)=2\sigma_\phi^2[1-C_\alpha(\tau;D)]$,
and $\sigma_\phi^2$ from Eq.~\eqref{eq:sigmaphi_cosmo}; at $\tau\to\infty$
($C_\alpha\to0$), $\mathcal{F}_{\rm foam}\to e^{-\sigma_\phi^2}$.

\subsection{Gaussian Source Profile}
\label{sec:closedforms}

The GRB\,221009A emission line is well described by a Gaussian profile
in each epoch~\cite{YQZhang2024}, and it serves as the representative
case admitting the simplest closed form.
A Gaussian line of $1\sigma$ width $\sigma_\nu$ centered on $\nu_0$
corresponds via the Wiener--Khinchin theorem to the coherence function
\begin{equation}
  g^{(1)}_{\rm src}(\tau) \;=\; e^{-i\omega_0\tau}\,e^{-\tau^2/2\tau_G^2},
  \label{eq:g1src_gauss}
\end{equation}
where $\omega_0 \equiv 2\pi\nu_0$ and $\tau_G \equiv 1/(2\pi\sigma_\nu)$ is the
Gaussian coherence decay time.
Substituting the Gaussian profile~\eqref{eq:g1src_gauss} into the
jitter-convolution integral~\eqref{eq:g1obs_carrier} and evaluating the
resulting Gaussian integral over $\Delta$ (with
$\sigma_\Delta^2\equiv\sigma_\Delta^2(\tau)$) gives the exact closed form
\begin{align}
  g^{(1)}_{\rm obs}(\tau)
  &= \frac{\tau_G}{\sqrt{\tau_G^2+\sigma_\Delta^2(\tau)}}\,
     \exp\!\left[-\,i\omega_0\tau
       +\frac{i\,\omega_0\,\sigma_\Delta^2(\tau)\,\tau}
              {\tau_G^2+\sigma_\Delta^2(\tau)}\right] \notag\\
  &\quad\times \exp\!\left[-\,\frac{\tau^2}
       {2\bigl[\tau_G^2+\sigma_\Delta^2(\tau)\bigr]}\right]
     \exp\!\left[-\,\frac{D_\phi(\tau)}{2}
       +\frac{\omega_0^2\sigma_\Delta^4(\tau)}
              {2\bigl[\tau_G^2+\sigma_\Delta^2(\tau)\bigr]}\right].
  \label{eq:g1obs_gauss}
\end{align}
The observed coherence is again a Gaussian-modulated oscillation, but
with three foam-induced modifications: (i) a slight shift of the carrier,
$\nu_0^{\rm eff} = \nu_0\bigl[1-
\sigma_\Delta^2(\tau)/(\tau_G^2+\sigma_\Delta^2(\tau))\bigr]$;
(ii) a wider Gaussian decay of $g^{(1)}_{\rm obs}(\tau)$ in lag (decay
scale $\sqrt{\tau_G^2+\sigma_\Delta^2(\tau)}$ vs.\ $\tau_G$);
and (iii) a global amplitude suppression set by the differential phase
variance $D_\phi(\tau)=\omega_0^2\sigma_\Delta^2(\tau)$ (Eq.~\eqref{eq:Ffoam}),
at the carrier $\omega_0$.  The marginal phase variance $D_\phi=\omega_0^2\sigma_\Delta^2$ is exact,
while the carrier shift~(i) and the $\omega_0^2\sigma_\Delta^4$ amplitude
term are carrier--envelope coupling of order $\sigma_\Delta^2/\tau_G^2$.
In the slowly-varying-envelope limit $\sigma_\Delta\ll\tau_G$ these vanish
and $\mathcal{F}_{\rm foam}=e^{-D_\phi/2}$ factors out cleanly,
$g^{(1)}_{\rm obs}\to\mathcal{F}_{\rm foam}(\tau)\,g^{(1)}_{\rm src}(\tau)$ ---
the pure multiplicative suppression on which the spectroscopic bounds rest.
That coupling is the finite-linewidth correction in disguise.  Because
$\sigma_\phi^2\propto\nu^2$, the weight $e^{-\sigma_\phi^2}$ is a per-mode
quantity, and a line of fractional width
$r=\sigma_\nu/\nu_0=1/(\omega_0\tau_G)$ does not carry a single one;
averaging it across the profile raises it by $\exp(2\sigma_\phi^4r^2)$, which
is exactly the $\omega_0^2\sigma_\Delta^4/(2\tau_G^2)$ above.
Equation~\eqref{eq:g1obs_gauss} therefore already carries the line average,
resummed into $\tau_G^2+\sigma_\Delta^2$.  It does not propagate to the
bounds in any case: the fractional width cancels the weight
(Section~\ref{sec:grb_width}), and the TeV and CMB probes apply
$e^{-\sigma_\phi^2(\nu)}$ frequency by frequency.
Normalization $g^{(1)}_{\rm obs}(0)=1$ holds because $\sigma_\Delta(0)=0$
(hence $D_\phi(0)=0$).
The narrowband limit $\tau_G\to\infty$ of
Eq.~\eqref{eq:g1obs_gauss} reduces to a pure phase-modulated
oscillation, which is the single-mode building block from which the
broadband CMB spectrum of Section~\ref{sec:CMB_derivation} is
constructed by incoherent superposition.

% ==============================================================
\section{Implications for Spectroscopy}
\label{sec:spec}
% ==============================================================

\subsection{The Wiener--Khinchin Bridge}

The foam-modified observed coherence function — given in general by the
jitter-convolution integral Eq.~\eqref{eq:g1obs} with the closed form
Eq.~\eqref{eq:g1obs_gauss} — carries direct
spectroscopic content: for a stationary field the power spectral density
$S(\nu)$ and the field autocorrelation
$\langle E^*(t)E(t+\tau)\rangle$ are Fourier-transform pairs,
implying
\begin{equation}
  S_{\rm obs}(\nu) = \langle|E_{\rm obs}|^2\rangle\,
  \mathcal{F}\!\left\{g^{(1)}_{\rm obs}(\tau)\right\}(\nu).
  \label{eq:WK_obs}
\end{equation}
Here $\mathcal{F}\{f\}(\nu)\equiv\int_{-\infty}^{\infty}d\tau\,
e^{\,i2\pi\nu\tau}f(\tau)$.  The theorem applies because
$E_{\rm src}$ and $\delta t$ are stationary and independent, and
the integration time exceeds the foam coherence time,
$T_{\rm int}\gg\tau_{\rm foam}$.  Since
$\tau_{\rm foam}\sim\tau_P\sim10^{-44}$\,s is far shorter than any
astrophysical evolution --- the $\sim\!10^2$\,s over which the GRB line
centroid sweeps $37\to6$\,MeV, say --- the slow kinematics and the fast foam jitter separate
cleanly, and each epoch is stationary on the foam timescale.  The prefactor
fixes the normalization: $g^{(1)}(0)=1$ gives
$\int S_{\rm obs}\,d\nu=\langle|E_{\rm obs}|^2\rangle$.

In the dynamical limit $C_\alpha(\tau;D)\to 0$ for
$|\tau|\gtrsim\tau_{\rm foam}$, $\mathcal{F}_{\rm foam}(\tau)$ acquires
a long-lag plateau $e^{-\sigma_\phi^2}$ outside a narrow window
$|\tau|\lesssim\tau_{\rm foam}$.  Fourier-transforming via
Eq.~\eqref{eq:WK_obs}, the plateau produces a coherent core
$e^{-\sigma_\phi^2}S_{\rm src}(\nu)$, while the short-lag window
redistributes the complementary fraction $1-e^{-\sigma_\phi^2}$ of the
source flux into a broad pedestal.  This core-plus-pedestal split holds
for any source, with $\sigma_\phi^2$ evaluated frequency by frequency
($\sigma_\phi^2\to\sigma_\phi^2(\nu)$) for a broadband source; its
broadband form for the CMB, including the pedestal
shape, is worked out for a specific model of the temporal
autocorrelation $C_\alpha$ in Section~\ref{sec:CMB_derivation}.
The core-plus-pedestal split is the optical analogue of the M\"ossbauer
recoil-free fraction: Gaussian phase statistics and unit modulus reduce any
such modulation to a sharp elastic line of weight $e^{-\sigma_\phi^2}$,
retaining the source profile, plus an inelastic remainder carrying
$1-e^{-\sigma_\phi^2}$ spread over the far broader bandwidth set by the
perturbation's correlation time.

Profiles in the remainder of this section are quoted per unit line
flux, $\int S\,d\nu=1$, so the core and pedestal weights read as fractions;
absolute units return in Section~\ref{sec:CMB_derivation}.
For the Gaussian source profile of Section~\ref{sec:closedforms},
Fourier-transforming Eq.~\eqref{eq:g1obs_gauss} with $\sigma_\Delta^2$
held at its asymptotic value $2\sigma_\ell^2/c^2$ gives the explicit
Gaussian core
\begin{equation}
  S_{\rm obs}^{\rm core}(\nu) \;=\;
    \frac{\tau_G}{\sqrt{\tau_G^2+\sigma_\Delta^2}}\,
    \exp\!\left[-\frac{\sigma_\phi^2\,\tau_G^2}{\tau_G^2+\sigma_\Delta^2}\right]
    \cdot
    \frac{1}{\sqrt{2\pi}\,\sigma_{\rm obs}}
    \exp\!\left[-\frac{(\nu-\nu_0^{\rm eff})^2}{2\sigma_{\rm obs}^2}\right],
  \label{eq:Sobs_gauss}
\end{equation}
with $\nu_0^{\rm eff}=\nu_0\tau_G^2/(\tau_G^2+\sigma_\Delta^2)$ and
$\sigma_{\rm obs}=1/[2\pi\sqrt{\tau_G^2+\sigma_\Delta^2}]<\sigma_\nu$; the
result is exact for the single-carrier Gaussian source of
Eq.~\eqref{eq:g1obs_gauss}.
The Gaussian's carrier shift, line narrowing, and peak suppression are
the frequency-domain images of items~(i)--(iii) of
Section~\ref{sec:closedforms}.

The complementary $1-e^{-\sigma_\phi^2}$ of the line flux reappears as a
pedestal: in the regime $\sigma_\Delta\ll\tau_G$, the plateau/residual
decomposition of $\mathcal{F}_{\rm foam}$ sends the short-lag residual to the
pedestal
\begin{align}
  S_{\rm obs}^{\rm ped}(\nu)
  &=\int_{-\infty}^{\infty}\!d\tau\,
    e^{\,i2\pi(\nu-\nu_0)\tau}\,e^{-\tau^2/2\tau_G^2}
    \left[e^{-\sigma_\phi^2[1-C_\alpha(\tau)]}-e^{-\sigma_\phi^2}\right] \notag\\
  &\xrightarrow{\;\tau_{\rm foam}\ll\tau_G\;}
    \hat{R}(\nu-\nu_0)=\tau_{\rm foam}\,\hat{h}\!\big[(\nu-\nu_0)\,\tau_{\rm foam}\big],
  \label{eq:Sobs_gauss_ped}
\end{align}
where the limit rescales $\tau=\tau_{\rm foam}x$, $h(x)\equiv
e^{-\sigma_\phi^2[1-C_\alpha(\tau_{\rm foam}x)]}-e^{-\sigma_\phi^2}$, and
$\hat{h}(y)\equiv\int_{-\infty}^{\infty}dx\,e^{\,i2\pi xy}\,h(x)$ is a fixed
$\mathcal{O}(1)$ function of unit-order width; the full observed line is
$S_{\rm obs}(\nu)=S_{\rm obs}^{\rm core}(\nu)+S_{\rm obs}^{\rm ped}(\nu)$.
The pedestal width $\Delta\nu_{\rm ped}\sim1/\tau_{\rm foam}$ exceeds the line
width $\sigma_\nu=1/(2\pi\tau_G)$ by $\tau_G/\tau_{\rm foam}\gg1$ (Fourier
reciprocity: narrow in lag, broad in frequency), and its total weight
$\int S_{\rm obs}^{\rm ped}\,d\nu=1-e^{-\sigma_\phi^2}$ complements the core's
$e^{-\sigma_\phi^2}$, conserving the integrated line flux.  Spread over
$\sim1/\tau_{\rm foam}$ --- for Planck-scale foam $\Delta\nu_{\rm ped}\sim10^{43}$\,Hz
--- the pedestal's in-line contribution is suppressed by $\tau_{\rm foam}/\tau_G\ll1$,
so the spectrum near the line peak is dominated by the coherent core.

The core and pedestal differ in predictive status.  The suppression
$e^{-\sigma_\phi^2(\nu)}$ is exact for any $\sigma_\phi^2$ --- from
Gaussian phase statistics and long-lag decorrelation ($C_\alpha\to0$) ---
so the linearization $1-e^{-\sigma_\phi^2}\simeq\sigma_\phi^2$ is only a
convenience.  The observed deficit equals $1-e^{-\sigma_\phi^2(\nu)}$
while the pedestal stays below the core.  Being broad
($\sim1/\tau_{\rm foam}$), the pedestal returns only a fraction
$f_{\rm band}\sim\Delta\nu\,\tau_{\rm foam}$ to a band $\Delta\nu$, so the
core exceeds this floor for $\sigma_\phi^2\lesssim-\ln f_{\rm band}$ ---
deep into the optically thick regime, the more so as self-convolution
broadens the pedestal.  Beyond it the deficit saturates at the floor:
optically thick foam remains evident, but its strength requires the
pedestal profile, the $C_\alpha$-dependent self-convolution series.
Flux conservation preserves the qualitative signature throughout.

% ==============================================================
\subsection{\texorpdfstring{Spectral Analogue of}{Spectral Analogue of} the Imaging Strehl Ratio}
\label{sec:connection}

The first-order coherence function $g^{(1)}_{\rm obs}(\tau)$ derived
above describes temporal coherence along a single line of
sight, with the foam suppression governed by the temporal
decorrelation $C_\alpha(\tau;D)$ of Eq.~\eqref{eq:Ffoam}.  The Strehl
ratio of an imaging telescope is the spatial counterpart:
phase variations across the aperture plane are governed by an
aperture-plane coherence function $C_\alpha^{\rm sp}(B;D)$, constructed
by direct analogy to the temporal $C_\alpha(\tau;D)$ of
Section~\ref{sec:Calpha_origin} under the same foam-cell statistics.

The structural equivalence is not merely formal: in both cases the
foam suppression arises as the characteristic function of a zero-mean
Gaussian differential random variable.  Temporally, the variable is
the differential jitter $\Delta(\tau)=\delta t(t+\tau)-\delta t(t)$
with variance $\sigma_\Delta^2(\tau)=(2\sigma_\ell^2/c^2)[1-C_\alpha(\tau;D)]$
(Eq.~\eqref{eq:diffVarExt}).  Spatially, the variable is the
differential phase
$\Delta\phi(B)=\phi(\mathbf{x}+\mathbf{B})-\phi(\mathbf{x})$
between two points on the aperture plane separated by baseline $B$,
with variance
\begin{equation}
  \sigma_{\Delta\phi}^2(B) \;=\; 2\sigma_\phi^2\bigl[1-C_\alpha^{\rm sp}(B;D)\bigr].
  \label{eq:dphi-variance}
\end{equation}

The focal-plane field of a telescope is the Fraunhofer Fourier
transform of the aperture-plane field, giving the diffraction-limited
PSF when the aperture phase is uniform.  Foam imprints a Gaussian
random phase $\phi(\mathbf{x})$ on the aperture with variance
$\sigma_\phi^2$ and two-point correlation $C_\alpha^{\rm sp}(B;D)$,
blurring this PSF.  In the resolved limit
$C_\alpha^{\rm sp}(D_{\rm tel};D)\to 0$ --- where aperture phases at
distinct points decorrelate --- the on-axis peak is attenuated by the
Strehl ratio
\begin{equation}
  S \;\simeq\; e^{-\sigma_\phi^2}
  \label{eq:strehl}
\end{equation}
(extended-Mar\'echal~\cite{NgPerlman2022}), the surrounding Airy
fringes are washed out by the same factor, and in the imaging
channel the complementary $1-e^{-\sigma_\phi^2}$ of the source flux is
scattered in angle into an incoherent halo --- the ``halo'' of optical
photometry.
The pedestal is its spectroscopic counterpart: the same fraction of
the flux, displaced in frequency rather than in angle; where the foam is
transversely uncorrelated across the aperture, the flux that leaves the image
core is the flux that leaves the coherent core
(Appendix~\ref{app:extended}).

Imaging and spectroscopic channels each probe $\sigma_\phi^2$ through
a Fourier transform --- aperture position $\leftrightarrow$ focal-plane
angle in imaging (Fraunhofer), lag $\leftrightarrow$ frequency in
spectroscopy (Wiener--Khinchin) --- with the rapid chromatic tightening
$\sigma_\phi\propto 1/\lambda$ exploited in both Ng \&
Perlman's~\cite{NgPerlman2022} Table~1 and the analyses of
Section~\ref{sec:grb221009a}.  The shared $\sigma_\phi^2$ that
suppresses the coherent diffraction-limited PSF in imaging also
suppresses the coherent line core in spectroscopy, placing the
published imaging
constraints~\cite{NgPerlman2022,LieuHillman2003,Christiansen2011,Perlman2015}
on a common footing with the present framework.

Two caveats temper the duality.  A quantitative cross-channel comparison
is model-dependent: it requires relating the spatial ($C_\alpha^{\rm sp}$)
and temporal ($C_\alpha$) foam correlation functions under a common
foam-cell model, so the constraints coincide only to the precision of
those shared details.  And the channels differ in their reference truth:
in imaging the aperture is known a priori, whereas in spectroscopy the
intrinsic source spectrum is itself uncertain, making the foam
contribution harder to isolate.

The two channels also differ in how much of that complementary
fraction a real instrument records.  The imaging literature builds competing
mappings from the variance relation of Eq.~\eqref{eq:NgPerlmanVar} to an
observable --- a Strehl ratio in Ref.~\cite{NgPerlman2022}, an angular size
subject to instrumental acceptance in
Refs.~\cite{Steinbring2007,Steinbring2015,SteinbringNg2026} --- and what
separates them is how much of the halo a telescope far from its diffraction
limit recovers, a fraction large enough that its treatment moves the inferred
$\alpha$.  That choice does not propagate here.  The factor
$e^{-\sigma_\phi^2}$ is derived in Section~\ref{sec:spec} from the
first-order coherence itself, and coincides with the Strehl factor only
because both are the Gaussian average $\langle e^{i\phi}\rangle$ --- taken
across the aperture there, across the lag here.  No angular inference enters,
so the bounds are unaffected by which mapping is correct; by the same token
the framework cannot be turned on \textit{Fermi}-LAT imaging data, which
record arrival directions rather than spectra.  An estimate of $\alpha$ from
those data would be an imaging analysis, made with a mapping this work does
not use.

The spectral counterpart of instrumental acceptance is the bandpass, and how
much of the pedestal is retained therein.  Because the pedestal is
$\sim1/\tau_{\rm foam}$ wide, that fraction
is $f_{\rm band}\approx4\times10^{-17}$ for LHAASO
(Section~\ref{sec:grb_tev}) and $\sim10^{-31}$ for FIRAS
(Section~\ref{sec:CMB_derivation}), so the displaced flux is effectively absent
from the measurement and the core deficit is observable without modeling its
return.  The instrumental line-spread function likewise does not enter: the
MeV line is resolved and its quoted widths are already response-corrected
(Section~\ref{sec:grb_width}); the TeV probe counts photons rather than
measuring a profile; and for FIRAS the foam deficit varies over the full
$60$--$600$\,GHz band, far broader than the $13.6$\,GHz channel spacing, with
the fit of Section~\ref{sec:CMB} carrying the apodized instrument line shape
through its covariance.  Angular scattering, whether or not it accompanies the
temporal jitter, is disposed of separately: the fractional width and the
monopole are insensitive to it, and Appendix~\ref{app:extended} shows the
focal-plane-integrated spectrum is independent of transverse foam structure
across the aperture.

% ==============================================================
\section{Sources}
\label{sec:sources}

Time-varying spacetime foam modifies the source first-order coherence,
and hence its spectrum, in a wavelength-dependent way through
the scaling $\sigma_\phi \propto 1/\lambda$. Three observable
consequences follow.  First, the coherent core loses absolute weight:
broadband flux is suppressed by $e^{-\sigma_\phi^2(\nu)}$ outright, whether or
not any line shape is resolved.  Second, spectral features distort per
Eq.~\eqref{eq:Sobs_gauss} with a carrier shift, line narrowing, and peak
suppression scaling as $\sigma_\phi\propto1/\lambda$.  Although the
absolute width narrows (item~(ii), Section~\ref{sec:closedforms}), the
center shifts to lower frequency by a larger factor (item~(i)), so the
fractional width $\sigma/\nu$ broadens --- the quantity
constrained below. Third, the
broadband spectral energy distribution is distorted: the cosmic
microwave background, expected to be a near-perfect blackbody, would
acquire foam-induced deviations whose chromatic signature is distinct
from standard chemical-potential and Compton-$y$ distortions.

\subsection{GRB\,221009A}
\label{sec:grb221009a}

\subsubsection{The source}
\label{sec:grb_source}

GRB\,221009A (``the BOAT,'' Brightest Of All Time) is the most
energetic gamma-ray burst in the instrumental record, with
isotropic-equivalent energy $E_{\rm iso}\sim10^{55}\,\mathrm{erg}$ at
redshift $z=0.151$~\cite{Ravasio2024}, a once-in-$\sim10^4$-year event~\cite{Burns2023}.
Adopting the Planck\,2018 flat $\Lambda$CDM cosmology used throughout~\cite{Aghanim2020}
($\Omega_m=0.315$, $\Omega_\Lambda=0.685$,
$H_0=67.36\,\mathrm{km\,s^{-1}\,Mpc^{-1}}$), the comoving distance is
$D_C(z=0.151)\approx0.65\,\mathrm{Gpc}\approx2.0\times10^{25}\,\mathrm{m}$
and $\ln(D_C/\ell_P)\approx138.4$.

Two features of this event make it a unique foam laboratory.
First, LHAASO detected more than $5000$ gamma rays above $0.5$\,TeV
(WCDA)~\cite{LHAASOWCDA2023} and $142$ events above $3$\,TeV (KM2A) extending to
$\sim\!13$\,TeV~\cite{LHAASOTeV2023}; these provide an absolute-flux
test of the foam suppression.  Second, two independent analyses report a
narrow gamma-ray emission feature in the Fermi/GBM data, interpreted as
blueshifted electron--positron pair
annihilation~\cite{Ravasio2024,YQZhang2024}, at observed energies
spanning $6$--$37$\,MeV.  Zhang et al.~\cite{YQZhang2024} detect a
Gaussian line whose centroid evolves as $E_{\rm line}\propto t^{-1}$
from $\approx37$\,MeV to $\approx6$\,MeV, with $t$ measured from the
fitted decay onset $t_0\approx226$\,s post-trigger, with a fractional width
$r_{\rm obs}\equiv\sigma_{\rm obs}/E_{\rm line}$ of weighted mean
$0.111\pm0.014$, the seven sub-intervals spanning $\approx0.07$--$0.20$ and
consistent with a constant across the $9$--$37$\,MeV lever arm
($\chi^2/\mathrm{dof}=5.3/6$; Table~\ref{tab:rwidth}); the foam analysis depends only
on the Doppler-invariance of the line's fractional width, not on the
emission mechanism.
The line is resolved: the BGO energy resolution near an MeV is
$\approx\!10\%$ FWHM, improving toward higher energies across the line's
$6$--$37$\,MeV span, so $\sigma_{\rm LSF}/E\lesssim0.04$, well below the
measured $r_{\rm obs}$, and the spectral fits of Ref.~\cite{YQZhang2024}
account for the instrument, so the widths quoted there are already
corrected for it.  The instrumental line-spread function therefore does not
enter the fractional-width bound.  Nor does it enter the TeV survival bound,
which counts photons rather than measuring a profile.

These two data sets yield two complementary foam probes that draw on the
same $\sigma_\phi^2\propto\nu^2$ physics but differ sharply in strength
and in what they assume.  The stronger is an absolute bound: the
mere survival of TeV photons bounds $\sigma_\phi^2$ directly.  The
weaker is a relative one: energy-dependent broadening of the MeV
line constrains $\alpha$ only in the deep strong-foam regime.  Both
operate in the dynamical-foam regime $\tau_{\rm foam}\ll\tau_c$ --- with
$\tau_c\equiv\tau_G=1/(2\pi\sigma_\nu)$ the coherence time of the light
(Eq.~\eqref{eq:g1src_gauss}) --- adopted
throughout Section~\ref{sec:model}; the opposite quasi-static limit is
addressed in Sections~\ref{sec:grb_tev} and ~\ref{sec:conclusions}.

\subsubsection{Survival of the LHAASO TeV photons}
\label{sec:grb_tev}

The foam-modified spectrum of a broadband source splits into a coherent
core and an incoherent pedestal,
\begin{equation}
  S_{\rm obs}(\nu)
    = e^{-\sigma_\phi^2(\nu)}\,S_{\rm src}(\nu)
    + \int S_{\rm src}(\nu')\,\hat{R}(\nu-\nu';\nu')\,d\nu',
  \label{eq:robs_decomp}
\end{equation}
where the first term is the surviving carrier,
suppressed per mode by $e^{-\sigma_\phi^2(\nu)}$, while the second is the
flux removed from the core, $1-e^{-\sigma_\phi^2}$, redistributed by the
pedestal kernel $\hat{R}$ of width $\sim1/\tau_{\rm foam}$.
Two assumptions govern whether the pedestal returns flux to the
observing band.

\paragraph{Assumption (i): dynamical foam.}  The pedestal exists only if foam
decorrelates, $C_\alpha(\tau)\to0$; the quasi-static limit $C_\alpha\to1$
yields $\mathcal{F}_{\rm foam}\to1$ with no suppression to constrain
(Section~\ref{sec:Calpha_origin}).  For Planck-time-refresh foam
($\tau_{\rm foam}\sim\tau_P\equiv\ell_P/c\approx5\times10^{-44}$\,s),
this escape is closed even at TeV: the margin
$\tau_c/\tau_{\rm foam}\gtrsim10^{22}$ at $37$\,MeV shrinks toward
higher energy as $\tau_c\propto1/\nu$ (with $\tau_c=1/(2\pi r_{\rm obs}\nu)$
at the constant fractional width $r_{\rm obs}=0.10$), but
remains $\gtrsim10^{17}$ at $3$\,TeV --- still enormous, so the foam is
dynamical throughout.
This energy-scaling argument does not address models with long-range
temporal correlations (the temporal analogue of
Ref.~\cite{LeeZurekChen2024}; see Section~\ref{sec:Calpha_origin}),
which give $C_\alpha\approx1$ regardless of photon energy and remain a
separate escape from all bounds in this paper.

\paragraph{Assumption (ii): bandwidth $\ll$ pedestal.}  The pedestal width is
$\sim c/\ell_P\approx10^{43}$\,Hz, whereas the entire electromagnetic
output of a GRB (radio to TeV) spans $\lesssim10^{27}$\,Hz, a ratio
$\sim10^{-16}$.  For a Lorentzian pedestal of half-width
$\Gamma=c/\ell_P\approx1.9\times10^{43}$\,Hz, the in-band kernel fraction
over the LHAASO acceptance ($3$--$13$\,TeV, $\Delta\nu\approx2.4\times10^{27}$\,Hz)
is $f_{\rm band}=\Delta\nu/(\pi\Gamma)\approx4\times10^{-17}$, so
scattered flux returning to band is negligible.  This evaluates the
pedestal at $\tau_{\rm foam}=\tau_P$; more generally the in-band fraction
is negligible only while the pedestal stays broader than the band,
$1/\tau_{\rm foam}\gg\Delta\nu_{\rm band}$, i.e.\
$\tau_{\rm foam}\lesssim1/\Delta\nu_{\rm band}\approx4\times10^{-28}$\,s.
This is far stronger than the dynamical-foam condition
$\tau_{\rm foam}\ll\tau_c\sim10^{-22}$\,s that the width bound requires: for
$10^{-28}\,{\rm s}\lesssim\tau_{\rm foam}\lesssim10^{-22}$\,s the foam is
still dynamical (the line-width bound holds) but the pedestal narrows into
the band and the TeV survival bound weakens.  The TeV exclusion of
$\alpha=2/3$ therefore rests specifically on near-Planck-refresh foam,
$\tau_{\rm foam}\lesssim4\times10^{-28}$\,s.

With the in-band pedestal negligible, the observed flux reduces to the
absolute suppression
\begin{equation}
  S_{\rm obs}(\nu)\Big|_{\rm in\text{-}band}
    \;\approx\; e^{-\sigma_\phi^2(\nu)}\,S_{\rm src}(\nu).
  \label{eq:absolute_suppression}
\end{equation}
This is an absolute attenuation, testable whenever the detected flux can
be compared to an independently predicted intrinsic flux.
Equation~\eqref{eq:absolute_suppression} holds while the surviving core
exceeds the pedestal floor, $\sigma_\phi^2\lesssim-\ln f_{\rm band}\approx38$
(Section~\ref{sec:spec}; distinct from the suppression strength
$\sigma_{\phi,\max}^2\sim1$--$10$ that the data actually constrain,
Eq.~\eqref{eq:tev_sigmaphi}); for stronger foam the in-band signal saturates at
the floor $\sim f_{\rm band}N_0$, an escape foreclosed below by the
transient, directional nature of the detection.

The LHAASO-KM2A detector registered $142$ gamma rays above $3$\,TeV
from GRB\,221009A at $20.6\,\sigma$ significance, with events extending
to $\sim\!13$\,TeV~\cite{LHAASOWCDA2023,LHAASOTeV2023}.  By
Eq.~\eqref{eq:absolute_suppression} the observed in-band count is
$N\approx e^{-\sigma_\phi^2(E_{\rm TeV})}\,N_0$, where
$N_0$ is the in-band count in the absence of foam (after absorption by the
extragalactic background light, EBL).
Detection of $N$ photons implies
\begin{equation}
  \sigma_\phi^2(E_{\rm TeV})\;\approx\;\ln\!\frac{N_0}{N}
  \;\equiv\;\sigma_{\phi,\max}^2,
  \label{eq:tev_sigmaphi}
\end{equation}
where $\sigma_{\phi,\max}^2$ is the number of $e$-foldings of
suppression the source can sustain before the foam-attenuated count
falls below the observed~$N$.
Writing $\sigma_\phi^2=(2\pi D_C/\lambda_{\rm obs})^2(\ell_P/D_C)^{2\alpha}$
from Eqs.~\eqref{eq:NgPerlmanVar} and \eqref{eq:sigmaphi_cosmo}
($A_\alpha=1$, $\lambda_{\rm obs}=hc/E_{\rm TeV}$) and inverting gives
\begin{equation}
  \alpha \;\gtrsim\;
  \frac{\ln\!\bigl(2\pi D_C/\lambda_{\rm obs}\bigr)
        -\tfrac12\ln\sigma_{\phi,\max}^2}
       {\ln(D_C/\ell_P)}
  \;\equiv\; \alpha_{\rm TeV}.
  \label{eq:alpha_grb_tev}
\end{equation}
The strength of this bound is best seen by reductio.  At the holographic
value $\alpha=2/3$,
Eqs.~\eqref{eq:NgPerlmanVar}--\eqref{eq:sigmaphi_cosmo} at
$E_{\rm TeV}=3$\,TeV give
$\sigma_\phi^2\approx7\times10^{8}$.  To produce the observed $142$
photons would therefore require an intrinsic in-band count
$N_0 = N\,e^{\sigma_\phi^2}\approx 142\,e^{7\times10^8}
\approx 10^{3\times10^8}$ --- exceeding any astrophysical source by
hundreds of millions of orders of magnitude.  The exclusion is immune to
any uncertainty in the intrinsic spectrum or EBL model.

The precise boundary depends on how many $e$-foldings the source can
tolerate: allowing $\sigma_{\phi,\max}^2=1$, $3$, or $10$ $e$-foldings
gives $\alpha_{\rm TeV}=0.740$, $0.736$, or $0.732$ respectively
($E_{\rm TeV}=3$\,TeV, $\ln(2\pi D_C/\lambda_{\rm obs})\approx102.4$,
$\ln(D_C/\ell_P)\approx138.4$).  At the highest detected energy
$E_{\rm TeV}\approx13$\,TeV the bound tightens to $\alpha\gtrsim0.75$.
The result is doubly logarithmic in $N_0/N$: a decade in the tolerated
suppression shifts $\alpha$ by less than $0.01$
(at $z=0.151$ the EBL optical depth at $3$\,TeV is only
$\tau_{\rm EBL}\sim0.3$).

The escape could instead lie in the pedestal kernel itself --- a
light-tailed $\hat{R}$ that, contrary to Assumption~(ii), returns a
non-negligible fraction of the redistributed flux to the observing band.
The transient, directional nature of the detection forecloses this: the
kernel acts on every source at all times, so appreciable band-refilling
would deposit flux as a persistent sky-wide background --- not as a
$20.6\,\sigma$ excess from the direction of GRB\,221009A, coincident
with the burst.
The $142$ TeV photons are therefore genuine survivors of the source's
beamed emission, and the suppression they constrain is real.  The same
reasoning bounds the kernel at the population level: a refilling kernel
would reprocess the entire steady $\gamma$-ray source population into a
diffuse glow far exceeding the observed background, an overproduction
sharpest in the $\gamma$-ray band itself because
$\sigma_\phi^2\propto\nu^2$ concentrates the reprocessing at high
energies.  That detecting the suppression requires a
once-in-$10^4$-year transient, rather than a uniformly glowing sky, is
itself the statement that the in-band refill is negligible.

Dynamical foam redistributes intensity in frequency.  The
modulation $e^{i\omega\delta t(t)}$ has unit modulus, so $|E_{\rm obs}(t)|^2
= |E_{\rm src}(t)|^2$ pointwise and the integrated flux $\int S\,d\nu$ is
conserved exactly, so the foam makes no net energy transfer from the metric;
what it changes is the distribution in $\nu$, the flux removed from the core
reappearing in the pedestal of width $\sim1/\tau_{\rm foam}$.  Because
$C_\alpha$ is the autocorrelation of a stationary process,
$\mathcal{F}_{\rm foam}$ is real and even in $\tau$ and the mean frequency of
the full spectrum is unchanged as well: the core redshift
$\nu_0^{\rm eff}<\nu_0$ is compensated by a blue-weighted pedestal.  Because flux conservation and the core-plus-pedestal split
follow from unit modulus alone, they survive the extreme limit
$\tau_{\rm foam}\to\tau_P$, where the pedestal width $1/\tau_{\rm foam}$
greatly exceeds the carrier $\omega_0$: that limit does not invalidate the
decomposition, but drives $\sigma_\phi^2$ into the regime where the
coherent core is exponentially suppressed.

This survival exclusion is consistent with the imaging-channel finding
that GRB\,221009A's localization, including the LHAASO TeV photons, admits
$\alpha=2/3$~\cite{SteinbringNg2026}: imaging constrains the spatial
foam correlations, which blur the wavefront but retain the photons, whereas
the survival bound constrains the temporal (dynamical) correlations,
which remove them from the band.  The two coincide only for static foam,
where neither produces a spectral effect.

\subsubsection{Energy-dependent line width}
\label{sec:grb_width}

The line-width probe is complementary because it is
relative: the absolute suppression $e^{-\sigma_\phi^2(\nu)}$
attenuates line peak and adjacent continuum identically, canceling in
the fractional width $\sigma/E$.  It therefore needs no absolute-flux
model --- foam is read off the line shape alone --- and is
pedestal-independent, immune to the redistribution physics
contested above, surviving rejection of Assumption~(ii), the only
escape from the TeV bound.  Its limitation is that, for a narrow line,
the fractional width responds to foam only at second order in the jitter,
so it constrains $\alpha$ only in the
strong-foam regime and yields a significantly weaker, if independent,
floor on~$\alpha$.

\paragraph{The intrinsic line as a Doppler invariant.}
The MeV feature is understood as a Doppler-shifted line from the relativistic
outflow --- most plausibly the blueshifted $511$\,keV electron--positron
annihilation line of pair-plasma clumps comoving with the
jet~\cite{Ravasio2024,YQZhang2024}, though what follows needs only that it is a
boosted line, not the specific transition.  Its observed centroid is then the
comoving line blueshifted by the bulk Doppler factor $\mathcal{D}$ of the
emitting material, $E_{\rm line}=\mathcal{D}\,m_ec^2/(1+z)$, so the centroid
measures $\mathcal{D}$ directly; the observed decline
$E_{\rm line}\propto t^{-1}$ from $\approx37$ to $\approx6$\,MeV
tracks the
falling jet Lorentz factor ($\Gamma\propto t^{-1}$, $\Gamma\approx84$ at
$37$\,MeV in the Zhang et al.\ modeling, viewed near $\theta\sim1/\Gamma$ so
that $\mathcal{D}\approx\Gamma$) as the outflow
evolves~\cite{YQZhang2024}.  Because a bulk boost multiplies every photon
energy --- line center and line width alike --- by the same $\mathcal{D}$, the
fractional width $r_{\rm int}=\sigma'/E'$ is boost-invariant: it equals
its comoving value at every epoch, independent of $\mathcal{D}$, and is
therefore expected to stay constant as the burst evolves and the centroid
sweeps downward.  This is what is observed --- $r_{\rm obs}=0.10\pm0.01$ across
the seven sub-intervals (Section~\ref{sec:grb_source}) --- so in the
conventional picture $r_{\rm int}$ is a fixed, epoch-independent property of the
source, set by the comoving line shape (the spread of clump Doppler factors and
the pair-plasma temperature~\cite{YQZhang2024}) and carrying no foam.
Quantum-gravity jitter, if present, imprints an additional,
energy-dependent broadening on top of this constant baseline, computed next.

\paragraph{Derivation of fractional-width broadening.}
The width law below follows from the slowly-varying-envelope limit
$\sigma_\Delta\ll\tau_G$, in which the foam factor multiplies the line
cleanly.
Eq.~\eqref{eq:Sobs_gauss} gives an observed core whose absolute width
$\sigma_{\rm obs}=1/[2\pi\sqrt{\tau_G^2+\sigma_\Delta^2}]$ is
narrower than the intrinsic $\sigma_\nu=1/(2\pi\tau_G)$
(item~(ii), Section~\ref{sec:closedforms}); however, the line center
shifts down to $\nu_0^{\rm eff}=\nu_0\tau_G^2/(\tau_G^2+\sigma_\Delta^2)$
(item~(i)) by a larger relative amount.
The fractional width is
$r_{\rm obs}\equiv\sigma_{\rm obs}/\nu_0^{\rm eff}$, with
$\nu_0^{\rm eff}$ the observed centroid.  Combining gives
\begin{equation}
  r_{\rm obs} = \frac{\sigma_{\rm obs}}{\nu_0^{\rm eff}}
  = \frac{1/(2\pi\sqrt{\tau_G^2+\sigma_\Delta^2})}
         {\nu_0\,\tau_G^2/(\tau_G^2+\sigma_\Delta^2)}
  = \frac{1}{2\pi\nu_0\tau_G}\,\frac{\sqrt{\tau_G^2+\sigma_\Delta^2}}{\tau_G}
  = r_{\rm int}\sqrt{1+\sigma_\Delta^2/\tau_G^2},
\end{equation}
so the fractional width broadens:
\begin{equation}
  r_{\rm obs}^2(E) \;=\; r_{\rm int}^2\bigl[1+\sigma_\Delta^2/\tau_G^2(E)\bigr]
                  \;=\; r_{\rm int}^2 + r_{\rm foam}^2(E),
  \label{eq:robs_decomp_width}
\end{equation}
with $r_{\rm foam}^2(E)\equiv r_{\rm int}^2\,\sigma_\Delta^2/\tau_G^2(E)$.
The line is a superposition over clumps with a spread of Doppler
factors (Section~\ref{sec:grb_source}), and Eq.~\eqref{eq:g1obs_gauss} is
nonlinear in $\tau_G$ through its prefactor and its carrier shift, so an
intensity-weighted mean over clumps is not that closed form at a mean
$\tau_G$; it is applied to the line as a whole, with $r_{\rm int}$ the
fractional width of the composite.

In the dynamical limit $\sigma_\Delta=\sqrt2\,\sigma_\ell/c$ saturates,
and with $\tau_G(E)=\lambda_{\rm obs}(E)/(2\pi c\,r_{\rm int})$ the foam
excess becomes
\begin{equation}
  r_{\rm foam}(E)
  \;=\; \sqrt{2}\,r_{\rm int}^2\,\sigma_\phi(E)
  \;=\; r_{\rm int}^2\cdot\frac{2\sqrt{2}\,\pi\,\sigma_\ell}{\lambda_{\rm obs}(E)}
  \;\propto\; E,
  \label{eq:rfoam_dyn}
\end{equation}
predicting chromatic broadening that grows linearly with photon energy.
The $r_{\rm int}^2$ prefactor makes the fractional width a
second-order foam probe: since
$\sigma_\Delta^2/\tau_G^2=2r_{\rm int}^2\sigma_\phi^2$, an order-unity
broadening of a narrow line ($r_{\rm int}\approx0.1$) requires
$\sigma_\phi^2\gtrsim40$.  At the holographic value $\alpha=2/3$ this
threshold is reached at $E_{\rm obs}\approx720$\,MeV (where
$\sigma_\phi^2=40$); a confirmed narrow line at or above this energy in a
$z\sim0.15$ GRB would exclude $\alpha=2/3$ through the width alone,
independent of the pedestal kernel.  Like the width bound itself, this
test operates in the strong-foam regime where the coherent core is
exponentially suppressed; it asks whether the line survives as a coherent
feature at all.

\begin{figure}[!htb]
\centering
\includegraphics[width=0.9\columnwidth]{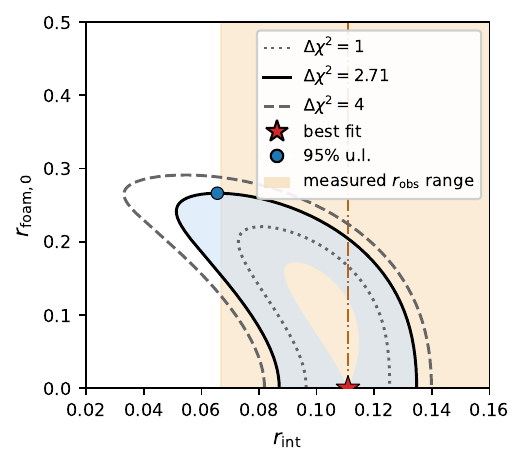}
\caption{$\Delta\chi^2$ in the $(r_{\rm int},r_{\rm foam,0})$ plane for the
line-width fit (contours at $1$, $2.71$, $4$). The best fit lies at the
boundary $r_{\rm foam,0}=0$, $r_{\rm int}=0.111$ (star); the one-sided
$95\%$ upper limit $r_{\rm foam,0}^{\rm max}=0.27$ is reached on the
$\Delta\chi^2=2.71$ contour at the profiled $r_{\rm int}=0.066$ (circle).
The shaded band marks the measured range of $r_{\rm obs}$
(Table~\ref{tab:rwidth}); the limit trades a narrower intrinsic line against
larger foam amplitude.}
\label{fig:grbchi2}
\end{figure}

\paragraph{Fit and bound.}
A $\chi^2$ fit of the seven measurements to the two-parameter model
\begin{equation}
  r_{\rm obs}^2(E) \;=\; r_{\rm int}^2 + r_{\rm foam,0}^2\,(E/E_0)^2,
  \qquad E_0=37\,\mathrm{MeV},
  \label{eq:robs_model}
\end{equation}
constrains the foam amplitude.  The best fit lies at the boundary
$r_{\rm foam,0}=0$ (with $r_{\rm int}=0.111$, $\chi^2_{\rm min}=5.3/6$\,dof),
confirming the constant-$r_{\rm obs}$ hypothesis.  Profiling over
$r_{\rm int}$ gives the $95\%$\,CL upper limit
$r_{\rm foam,0}^{\rm max}=0.27$ at $\Delta\chi^2=2.71$ (Fig.~\ref{fig:grbchi2}; boundary-corrected
Chernoff threshold~\cite{Chernoff1954}: $\tfrac12\delta_0+\tfrac12\chi_1^2$
for zero-bounded parameters; the precise threshold barely affects $\alpha$,
the logarithmic lever arm dominating).  Converting
via Eq.~\eqref{eq:rfoam_dyn} and $\sigma_\ell^2=D_C^{2(1-\alpha)}\ell_P^{2\alpha}$
(Eq.~\eqref{eq:NgPerlmanVar}, $A_\alpha=1$) gives
\begin{equation}
  \alpha \;\gtrsim\;
  \frac{\ln(D_C/\lambda_{\rm obs}) - \ln r_{\rm foam,0}^{\rm max}
        + \ln(2\sqrt{2}\,\pi\,r_{\rm int}^2)}
       {\ln(D_C/\ell_P)}
  \;=\; \frac{89.3 + 1.3 - 3.3}{138.4}
  \;\approx\; 0.63\,.
  \label{eq:alpha_grb_dyn}
\end{equation}
Here $r_{\rm int}=0.066$ is the width profiled jointly with
$r_{\rm foam,0}^{\rm max}$ on the $95\%$ boundary (not the global best fit
$0.111$), giving $|\!\ln(2\sqrt2\,\pi\,r_{\rm int}^2)|\approx3.3$.  Like
the TeV bound, the result is doubly logarithmic and so dominated by the
kinematic lever arm $\ln(D_C/\ell_P)=138$: a tenfold change in
$r_{\rm foam,0}^{\rm max}$ moves $\alpha$ by only
$\ln10/\ln(D_C/\ell_P)\approx0.017$, so $\alpha\gtrsim0.63$ is fixed by the
geometry, not by the precision of the seven measurements.

\begin{table}[!htb]
\caption{The seven Zhang et al.~\cite{YQZhang2024} MeV-line sub-intervals
used in the line-width fit: post-trigger interval, line energy $E$, observed
fractional width $r_{\rm obs}=\sigma_{\rm obs}/E$, and its uncertainty
$\sigma_r$. The weighted mean is $r_{\rm obs}=0.111\pm0.014$; the values are
consistent with a constant ($\chi^2/\nu=5.3/6=0.88$).}
\label{tab:rwidth}
\begin{ruledtabular}
\begin{tabular}{lccc}
interval [s] & $E$ [MeV] & $r_{\rm obs}$ & $\sigma_r$ \\
\hline
$246$--$256$ & $37.2$ & $0.191$ & $0.088$ \\
$275$--$280$ & $17.8$ & $0.093$ & $0.041$ \\
$285$--$290$ & $15.0$ & $0.129$ & $0.046$ \\
$290$--$295$ & $12.1$ & $0.120$ & $0.030$ \\
$295$--$300$ & $12.5$ & $0.067$ & $0.031$ \\
$300$--$310$ & $8.9$  & $0.196$ & $0.061$ \\
$310$--$320$ & $9.7$  & $0.115$ & $0.030$ \\
\end{tabular}
\end{ruledtabular}
\end{table}

\paragraph{Validity and caveats.}
At the $95\%$ boundary $\sigma_\Delta/\tau_G(E_0)\approx4.1$, i.e.\
$\sigma_\phi^2(37\,{\rm MeV})\sim10^3$, the bound lies deep in the
strong-foam regime, and three caveats attach to it.  First, the
fractional-width law Eq.~\eqref{eq:robs_decomp_width} is a property of the
coherent core, whose weight $e^{-\sigma_\phi^2}$ is exponentially small
there; the bound thus extrapolates the core's width relation into a regime
where the line flux has largely been redistributed into the pedestal, and
is best read as an order-of-magnitude exclusion anchored by the logarithmic
lever arm rather than a sharp threshold.  Second, the same
Eq.~\eqref{eq:g1obs_gauss} predicts a centroid depression
$\nu_0^{\rm eff}/\nu_0=\tau_G^2/(\tau_G^2+\sigma_\Delta^2)\approx0.06$ at
$37$\,MeV at this boundary, absent from the smooth
$E_{\rm line}\propto t^{-1}$ trend in the data; the width-only fit treats
observed centroids as intrinsic because a foam redshift is degenerate with
the kinematic $E_{\rm line}\propto t^{-1}$ evolution the data already
exhibit.  The centroid is the first-order, in principle stronger, foam
probe, so the second-order width bound is conservative; a joint
centroid-plus-width analysis is left to future work.  Third, the bound rests on the seven sub-interval
widths of the single GRB\,221009A feature, whose most constraining energy
($37$\,MeV) lies in the Fermi MeV sensitivity gap
(Section~\ref{sec:conclusions}), and the constant-$r_{\rm int}$ fit cannot
distinguish a genuine null from small epoch-to-epoch variation of the
intrinsic width.  With these caveats, the width-only $\alpha\gtrsim0.63$
stands as the pedestal-independent floor beneath the TeV bound.

% ==============================================================
\subsection{CMB}
\label{sec:CMB}

The CMB blackbody is itself a broad, asymmetric spectral profile, so foam
acts on it exactly as on any line: the coherent core is suppressed by
$e^{-\sigma_\phi^2(\nu)}$ and the removed flux reappears as a broad
pedestal whose shape depends on the short-time behavior of
$C_\alpha(\tau)$.  Because
$\sigma_\phi^2\propto\nu^2$ (Eq.~\eqref{eq:sigmaphi_cosmo}), the foam deficit
is concentrated at high frequencies --- a $\nu^2$-weighted suppression
distinct from the standard chemical-potential ($\mu$) and Compton-$y$
distortions.  For Planck-scale foam the pedestal is so broad that it lies
almost entirely outside the observing band (Section~\ref{sec:CMB_derivation}),
leaving an essentially pure deficit in the FIRAS window.

The Far-InfraRed Absolute Spectrophotometer (FIRAS) aboard the Cosmic
Background Explorer (COBE) measured the absolute spectrum of the CMB
monopole over $60$--$600$\,GHz by direct comparison against an internal
cryogenic blackbody calibrator --- the absolute-photometry feature that
distinguishes it from relative-photometry experiments and makes it
well suited to constraining spectral shape distortions.  The
calibrated monopole agrees with a pure blackbody to within
$\sim 50$\,ppm (a few$\,\times10^{-5}$ fractional) across the band,
placing the still-standard limits $|\mu|\lesssim 9\times10^{-5}$ and
$|y|\lesssim 1.5\times10^{-5}$ \cite{Fixsen1996}.  These limits and the
$43$-channel calibrated spectrum form the empirical basis for the foam
constraint derived below.

\subsubsection{Foam phase variance at CMB frequencies}
\label{sec:CMB_phasevar}

CMB photons travel from the last-scattering surface at comoving distance
$D_{\rm C}= 13.87\,{\rm Gpc}$ (Planck\,2018 flat-$\Lambda$CDM cosmology~\cite{Aghanim2020};
$H_0 = 67.36$\,km\,s$^{-1}$\,Mpc$^{-1}$, $\Omega_m=0.315$).  The foam phase
variance (Eq.~\eqref{eq:sigmaphi_cosmo}) is set by the effective foam
amplitude $A_{\rm eff}=A_\alpha D_{\rm C}^{2(1-\alpha)}\ell_P^{2\alpha}$
(the lumped amplitude entering the FIRAS fit,
Section~\ref{sec:CMB_degeneracy}):
\begin{equation}
  \sigma_\phi^2(\nu)
    = \left(\frac{2\pi\nu}{c}\right)^{\!2}\! A_{\rm eff},
  \label{eq:sigma_phi_CMB}
\end{equation}
with $A_\alpha=1$.

At $\nu=150$\,GHz this gives
\begin{equation}
  \sigma_\phi^2(150\,{\rm GHz})\approx
  \begin{cases}
    7\times10^{-2}  & \alpha=1/2\ \text{(random walk)},\\[2pt]
    2\times10^{-22}      & \alpha=2/3\ \text{(holographic)}.
  \end{cases}
  \label{eq:sigma_phi_CMB_values}
\end{equation}
For $\alpha=1/2$ the variance reaches order unity at the top of the
FIRAS band ($\sigma_\phi^2\approx1.2$ at $639$\,GHz), and the linearised
template holds for $\alpha\gtrsim0.50$, where $\sigma_\phi^2(\nu)\lesssim1$
through the band (Section~\ref{sec:CMB_degeneracy}).

\subsubsection{Foam-modified CMB spectrum}
\label{sec:CMB_derivation}

Given a foam model, the natural analysis is of the whole modified
spectrum: compute $g^{(1)}_{\rm src}(\tau)$ as the normalized
Fourier transform of
$B_\nu(T_0)$ (Eq.~\eqref{eq:WK_obs}), apply the jitter
convolution~\eqref{eq:g1obs}, and inverse-transform to $S_{\rm obs}(\nu)$.
We instead separate it into a coherent core and a pedestal, because the
split isolates a model-independent piece --- the suppression
$e^{-\sigma_\phi^2(\nu)}$, exact for any $C_\alpha$
(Section~\ref{sec:spec}) --- from the model-dependent pedestal
kernel $\hat{R}(\nu-\nu';\nu')$, a decomposition that yields general
physical insight the combined spectrum obscures.  A worked example with
exponential $C_\alpha(\tau)$ below identifies the physical no-pedestal
case used in Section~\ref{sec:CMB_degeneracy}.

Decomposing the blackbody as an incoherent superposition of narrowband
modes $S_{\rm src}(\nu')=B_{\nu'}(T_0)$, each obeys the
same achromatic jitter with per-mode foam factor
\begin{equation}
  \mathcal{F}_{\rm foam}(\tau;\nu')
    = e^{-\sigma_\phi^2(\nu')[1-C_\alpha(\tau)]},
  \quad
  \sigma_\phi^2(\nu')=(2\pi\nu'/c)^2\sigma_\ell^2(D),
  \label{eq:Ffoam_CMB}
\end{equation}
with $\sigma_\ell^2(D)=A_{\rm eff}$ (Section~\ref{sec:CMB_phasevar}).  By
the core-plus-pedestal split of Section~\ref{sec:spec}, in the dynamical
limit each mode's long-lag plateau $e^{-\sigma_\phi^2(\nu')}$
Fourier-transforms to a coherent core at $\nu'$ and its short-lag
residual to a broad pedestal; superposing over the Planck modes
(intensity-weighted Eq.~\eqref{eq:WK_obs}) gives
\begin{equation}
  S_{\rm obs}(\nu)
    = e^{-\sigma_\phi^2(\nu)}\,B_\nu(T_0)
    + \int_0^\infty B_{\nu'}(T_0)\,\hat{R}(\nu-\nu';\nu')\,d\nu',
  \label{eq:CMB_spectrum}
\end{equation}
with $\hat{R}$ the Fourier transform of the residual, carrying the full
$C_\alpha$-dependence.  By the unitarity of the per-mode modulation
$e^{i\omega\delta t(t)}$, the pedestal conserves the integrated
flux, $\int[S_{\rm obs}(\nu)-B_\nu(T_0)]\,d\nu=0$, the displaced flux
emerging at $\nu\sim1/\tau_{\rm foam}$.

To make the pedestal kernel concrete, we evaluate $\hat{R}$ for the
minimal model $C_\alpha(\tau)=e^{-|\tau|/\tau_{\rm foam}}$ (a Markovian
metric fluctuation).  Substituting into Eq.~\eqref{eq:diffVarExt} gives
\begin{equation}
  \sigma_\Delta^2(\tau) = \frac{2\sigma_\ell^2(D)}{c^2}
    \bigl[1 - e^{-|\tau|/\tau_{\rm foam}}\bigr].
  \label{eq:diffVarExt_exponential}
\end{equation}
The Fisher analysis below operates in the linearised regime
$\sigma_\phi^2(\nu')\ll 1$.  Expanding $\mathcal{F}_{\rm foam}(\tau;\nu')$
(Eq.~\eqref{eq:Ffoam_CMB}) to first order and subtracting the long-lag asymptote
$e^{-\sigma_\phi^2(\nu')}$ leaves the short-$\tau$ residual
\begin{equation}
  \mathcal{F}_{\rm foam}(\tau;\nu') - e^{-\sigma_\phi^2(\nu')}
  \;\approx\; \sigma_\phi^2(\nu')\,e^{-|\tau|/\tau_{\rm foam}},
  \label{eq:Ffoam_residual_exponential}
\end{equation}
whose Fourier transform is a Lorentzian centered on the source
frequency,
\begin{equation}
  \hat{R}(\nu-\nu';\nu')
    \;=\; \sigma_\phi^2(\nu')\,
    \frac{2\tau_{\rm foam}}{1 + [2\pi(\nu-\nu')\tau_{\rm foam}]^2},
  \label{eq:R_lorentzian}
\end{equation}
of full width at half maximum $1/(\pi\tau_{\rm foam})$.  For Planck-scale
foam with $\tau_{\rm foam}\sim\ell_P/c\approx 5\times 10^{-44}$\,s, the
Lorentzian width ($\sim 10^{43}$\,Hz) lies thirty-one orders of magnitude
beyond the FIRAS upper edge ($\nu_{\rm max}=6\times 10^{11}$\,Hz), so the
in-band flux fraction is $\sim 10^{-31}$.  The physical pedestal thus
deposits no detectable in-band power for any plausible $\tau_{\rm foam}$.  The
constraint of Section~\ref{sec:CMB_degeneracy} consequently
adopts the no-pedestal limit as the primary template.

\subsubsection{Constraint from FIRAS}
\label{sec:CMB_degeneracy}

We use $I_\nu\equiv S_{\rm obs}(\nu)$ to denote the observed CMB
intensity including the foam signal.
Working in the linearised regime
$\sigma_\phi^2(\nu)\ll 1$, we treat four spectral parameters
$\boldsymbol{\theta}=(A_{\rm eff},T_0,\mu,y)$, where
$A_{\rm eff}=A_\alpha D_C^{2(1-\alpha)}\ell_P^{2\alpha}$
is the cumulative comoving foam variance at the CMB distance
(Section~\ref{sec:CMB_phasevar}), so that
$\sigma_\phi^2(\nu)=(2\pi\nu/c)^2 A_{\rm eff}$.  Note that $\alpha$ is
not fitted directly; only the lumped amplitude $A_{\rm eff}$ enters
the fit, and bounds on $\alpha$ are obtained
\emph{a posteriori} by inverting the foam relation at fixed
$A_\alpha=1$.

For physical foam the pedestal contributes no in-band power
(Section~\ref{sec:CMB_derivation}), so the only in-band foam signature is
the pure deficit --- the no-pedestal limit of Eq.~\eqref{eq:CMB_spectrum} ---
\begin{equation}
  \frac{\partial I_\nu}{\partial A_{\rm eff}}
    = -\!\left(\frac{2\pi\nu}{c}\right)^{\!2} B_\nu(T_0),
  \label{eq:dIdA_phys}
\end{equation}
a $\nu^2$-weighted (blue-tilted) Planck spectrum.  The
chemical-potential and Compton-$y$ distortion templates are
\begin{align}
  \frac{\partial I_\nu}{\partial\mu}
    &= -B_\nu(T_0)\,\frac{x e^x}{(e^x-1)^2}, \label{eq:dIdmu}\\
  \frac{\partial I_\nu}{\partial y}
    &= \phantom{-}B_\nu(T_0)\,\frac{x e^x}{e^x-1}
       \!\left[\frac{x(e^x+1)}{e^x-1}-4\right], \label{eq:dIdy}
\end{align}
with $x=h\nu/(k_B T_0)$, encoding respectively early-universe energy
injection (Silk damping, decaying particles) and late-time electron
heating (reionization, cluster gas).

We constrain $A_{\rm eff}$ by fitting these templates --- together with
the FIRAS Galactic-foreground template --- to the public COBE/FIRAS
monopole residual spectrum and its calibration
covariance~\cite{Fixsen1996}.  Raising $T_0$ brightens the Planck curve
broadband; the foam deficit suppresses it with an extra $\nu^2$ tilt.
Over the CMB-bright band (peak through the onset of the Wien fall-off),
the two differ only by this tilt, making the foam template nearly
indistinguishable from a temperature shift
($\rho_{A_{\rm eff}T_0}\simeq 1$ in the fit) and thus strongly collinear
with the $y$ and $\mu$ templates.  Using the
full $43\times43$ apodization covariance, the generalized
least-squares fit over $(T_0,\mu,y,A_{\rm eff})$ and the foreground
amplitude recovers $\mu$ and $y$ consistent with zero (foam-marginalized
$95\%$ limits $|\mu|<1.7\times10^{-4}$, $|y|<1.0\times10^{-5}$;
$\chi^2/\nu=1.27$ --- the $\mu$ limit looser than the foam-free Fixsen
value $|\mu|\lesssim9\times10^{-5}$ because marginalizing over the foam
amplitude inflates $\sigma(\mu)$) and a foam
amplitude consistent with zero,
\begin{equation}
  A_{\rm eff}=(-2.4\pm5.3)\times10^{-11}\,{\rm m}^2 .
  \label{eq:Aeff_firas}
\end{equation}
Inverting $A_{\rm eff}=A_\alpha D_C^{2(1-\alpha)}\ell_P^{2\alpha}$
at $A_\alpha=1$, the one-sided $95\%$ upper limit (measured from zero, as
required by the physical boundary $A_{\rm eff}\geq0$)
$A_{\rm eff}<1.645\,\sigma(A_{\rm eff})=8.7\times10^{-11}\,{\rm m}^2$ gives the lower bound
$\alpha\gtrsim 0.52$ (collected with the forecasts below
in Table~\ref{tab:thresholds}).  This bound is degeneracy-limited: $\alpha\gtrsim0.52$ rests only on the
residual spectral-tilt information that distinguishes the
$\nu^2$-weighted deficit from a temperature shift, so marginalizing over
$(T_0,\mu,y)$ leaves it weak.
The only direction that could break the
foam--$T_0$ degeneracy is deep into the Wien/sub-mm regime, where the
$\nu^2$-tilt separates from a temperature shift; but there, beyond the
FIRAS band, galactic dust and the cosmic infrared background dominate,
making the monopole impractical.  The fit uses the linearised template
$\Delta I_\nu\approx\sigma_\phi^2(\nu)\,[\text{template}]$ (valid for
$\sigma_\phi^2\ll1$), which holds across the FIRAS band at the headline
threshold: $\sigma_\phi^2$ exceeds unity only for stronger, excluded foam
($\alpha\lesssim0.50$; Section~\ref{sec:CMB_phasevar}).  Replacing the
linearised deficit $-\sigma_\phi^2 B_\nu$ by the exact form
$-(1-e^{-\sigma_\phi^2})B_\nu$ (valid at all $\sigma_\phi^2$,
Section~\ref{sec:spec}) shifts the integrated in-band amplitude by
$\approx13\%$ at $\alpha=0.50$ but by $<1\%$ at $\alpha\gtrsim0.52$
($\Delta\alpha\sim3\times10^{-5}$), so the bound is unaffected; in that
strong-foam regime nonlinearity may even help break the foam--$\mu$
degeneracy.

The foam--$T_0$ degeneracy has two consequences for the CMB
temperature itself.  First, admitting foam as a free parameter inflates the
marginalized temperature uncertainty: the FIRAS spectral-shape
determination of $T_0$ degrades from $1.2\times10^{-5}$\,K to
$7.7\times10^{-4}$\,K (a factor $\sim 64$; $\sigma(y)$ and
$\sigma(\mu)$ inflate by $\sim 17$ and $\sim 3$), comparable
to the full COBE/FIRAS calibration uncertainty
$5.7\times10^{-4}$\,K~\cite{Fixsen2009}.  A foam-agnostic monopole
analysis thus roughly doubles the temperature error budget, and the quoted
precision of $T_0=2.72548$\,K implicitly assumes no foam.  Second, foam
present but omitted from the fit shifts the central value by
$\Delta T_0 = b_{T_0}A_{\rm eff}$ with
$b_{T_0}\approx -1.5\times10^7\,{\rm K\,m^{-2}}$; for $A_\alpha=1$ this
is only $\approx 3\times10^{-22}$\,K at the holographic $\alpha=2/3$ and
vanishes as $\alpha\to1$ (where the comoving foam variance ceases to grow
with distance), reaching $\sigma(T_0)$ only
at $\alpha\approx 0.52$ where the foam is already a many-$\sigma$
signal.  Both effects therefore matter only for foam near FIRAS
detectability; the recovered $T_0$ is immune to any foam consistent
with the GRB bound $\alpha\gtrsim 0.63$.

\subsubsection{Fisher forecasts for future spectrometers}
\label{sec:CMB_forecast}

\begin{figure}[!htb]
\centering
\includegraphics[width=0.82\columnwidth]{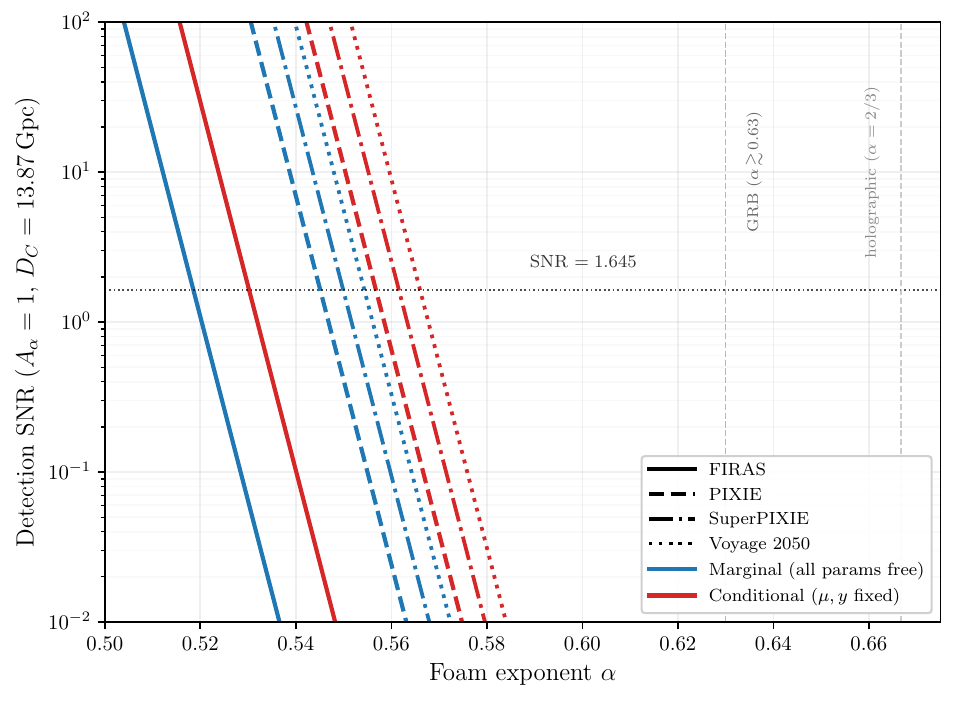}
\caption{Detection SNR for the foam amplitude $A_{\rm eff}$ versus
exponent $\alpha$ ($A_\alpha=1$, $D_C=13.87$\,Gpc).
Blue: marginal SNR (all four parameters free).
Red: conditional SNR ($\mu,y$ fixed), $\sim 27\times$ higher.
Line styles: solid FIRAS, dashed PIXIE, dash-dotted SuperPIXIE,
dotted Voyage~2050 (forecast $\mu$-sensitivities scaled to
FIRAS; \cite{Chluba2021}).
The marginal lower bounds cluster at $\alpha\approx 0.52$--$0.55$.
Horizontal dotted line: SNR\,$=1.645$ (one-sided $95\%$ CL).
Vertical dashed lines: the spectroscopic lower bound from
GRB\,221009A ($\alpha\gtrsim 0.63$) and the holographic prediction
$\alpha=2/3$.}
\label{fig:cmb_snr}
\end{figure}

To forecast future absolute spectrometers we propagate the no-pedestal
templates through a Fisher matrix with flat per-channel (white) noise
$\sigma_\nu=0.01$\,MJy/sr calibrated to FIRAS (the full $43\times43$
apodization covariance enters the data fit of
Section~\ref{sec:CMB_degeneracy}).  The forecast gives
$\sigma(A_{\rm eff})=2.2\times10^{-11}\,{\rm m}^2$ and
$\alpha\gtrsim 0.52$ (reproducing the data fit); the conditional value
$\sigma(A_{\rm eff})|_{\mu,y\,\text{fixed}}=8.2\times10^{-13}\,{\rm
m}^2$ shows that marginalizing over distortions costs a factor
$\sim 27$ via the foam--$T_0$ collinearity.  The fiducial
degeneracy structure is captured by the correlation matrix for
$(A_{\rm eff},\,T_0,\,\mu,\,y)$,
\begin{equation}
\rho^{\rm fid}_{ij} = \begin{pmatrix}
  1.000 & \mathbf{0.9997} & 0.928 & 0.999 \\
  \mathbf{0.9997} & 1.000 & 0.934 & 0.998 \\
  0.928 & 0.934 & 1.000 & 0.915 \\
  0.999 & 0.998 & 0.915 & 1.000
\end{pmatrix},
\label{eq:fisher_corr_matrix_fid}
\end{equation}
the foam deficit is almost perfectly aligned with a
temperature shift ($\rho_{A_{\rm eff}T_0}=0.9997$, bold) and strongly
collinear with the $y$ and $\mu$ distortions ($0.999$ and $0.928$).
Defining the marginal detection SNR as $A_{\rm eff}/\sigma(A_{\rm
eff})$, so that ${\rm SNR}=1.645$ marks the one-sided $95\%$ CL threshold,
Figure~\ref{fig:cmb_snr} plots it versus $\alpha$, and
Table~\ref{tab:thresholds} collects the thresholds.

\begin{table}[!htb]
\caption{Marginal foam-detection thresholds (one-sided $95\%$ CL).
Lower bounds on $\alpha$ ($A_\alpha=1$): the foam signal
$\sigma_\ell^2\propto(\ell_P/D_C)^{2\alpha}$ decreases with $\alpha$, so
non-detection excludes small $\alpha$.  FIRAS: data fit
(Section~\ref{sec:CMB_degeneracy}).  PIXIE, SuperPIXIE, Voyage~2050:
Fisher forecast scaled to published $\mu$-sensitivities~\cite{Chluba2021}.
The three forecast entries ($0.545$, $0.549$, $0.554$) round to
$0.54$--$0.55$.}
\label{tab:thresholds}
\begin{ruledtabular}
\begin{tabular}{lc}
 & lower bound \\
\hline
FIRAS        & $\alpha\gtrsim 0.52$ \\
PIXIE        & $\alpha\gtrsim 0.54$ \\
SuperPIXIE   & $\alpha\gtrsim 0.55$ \\
Voyage~2050  & $\alpha\gtrsim 0.55$ \\
\end{tabular}
\end{ruledtabular}
\end{table}

The no-pedestal limit is also the conservative one.  The pure
$\nu^2$-deficit is already near-perfectly aligned with a temperature
shift ($\rho_{A_{\rm eff}T_0}=0.9997$); any in-band pedestal redistributes
power back into the band, moving the foam template off that alignment, so
it can only tighten the bound.  A $\mu$-shaped kernel, for
instance, trades the near-perfect $T_0$ degeneracy for a weaker
foam--$\mu$ one ($0.976$), nudging the threshold to $\alpha\gtrsim0.53$.
This is moot for physical foam, whose pedestal is negligible in-band for
any $C_\alpha$ with $\tau_{\rm foam}\ll1/\nu_{\rm max}$
(Section~\ref{sec:CMB_derivation}); an external $\mu$ prior likewise
barely helps, since the limiting degeneracy is with $T_0$, not $\mu$.

The proposed PIXIE satellite~\cite{Kogut2011} --- a fully cryogenic
($T_{\rm sys}\approx2.7$\,K) nulling polarizing Fourier transform spectrometer
spanning $30$\,GHz to $6$\,THz --- is projected to surpass FIRAS sensitivity
by roughly three orders of magnitude, reaching $|\mu|\lesssim10^{-8}$ and
$|y|\lesssim10^{-9}$ and entering the regime of guaranteed $\Lambda$CDM
spectral signals from Silk damping and reionization.
Propagating this gain through the Fisher-matrix scaling derived above,
the marginal foam detection threshold shifts from
$\alpha\gtrsim0.52$ (FIRAS) to approximately $\alpha\gtrsim0.54$
(PIXIE; Table~\ref{tab:thresholds}), narrowing the gap with the
spectroscopic lower bound $\alpha\gtrsim0.63$ from GRB\,221009A
(Fig.~\ref{fig:cmb_snr}).
The foam--$T_0$ degeneracy persists on the linearized template.  But
PIXIE's extension to $6$\,THz enters the nonlinear regime
$\sigma_\phi^2\gg1$ for $\alpha\lesssim0.52$ (higher than the FIRAS
crossing $\alpha\lesssim0.50$ because the wider band pushes the
$\sigma_\phi^2=1$ boundary upward; Section~\ref{sec:CMB_degeneracy}),
where the foam deficit diverges markedly from the blackbody and
$\mu$-distortion shapes, possibly reducing the degeneracy.  The
$\alpha\gtrsim0.54$ projection may therefore understate the reach: if
the nonlinear template effectively breaks this degeneracy at PIXIE
sensitivity, the true threshold could be significantly more stringent.

Extending the same Fisher analysis to the $10$\,GHz--$3$\,THz bands of
the SuperPIXIE and Voyage~2050 concepts~\cite{Chluba2021} (forecast
distortion sensitivities $\sigma(\mu)\approx 7.7\times10^{-9}$ and
$2.2\times10^{-9}$) shifts the marginal threshold only to $\alpha\gtrsim 0.55$
(Table~\ref{tab:thresholds}), barely beyond the
PIXIE projection.  The reason is structural rather than instrumental:
the foam deficit $-(2\pi\nu/c)^2 B_\nu$ and the temperature derivative are
nearly proportional across the band, so the foam--$T_0$ degeneracy is set
by the templates, not the noise weighting.  Repeating the Fisher analysis
with the full $43\times43$ FIRAS apodization covariance rather than the
white-noise forecast gives $\rho_{A_{\rm eff}T_0}=0.9999$ (and a
marginal-to-conditional inflation $\approx25$) --- the degeneracy is thus
marginally more severe under realistic correlated noise, so the white-noise
estimate is conservative.  The
wider band does break the foam--$\mu$ degeneracy that an in-band
$\mu$-shaped pedestal would exploit (from
$\rho_{A_{\rm eff}\mu}=0.98$ at FIRAS to $0.66$ over
$10$\,GHz--$3$\,THz), so future missions are markedly more robust
against in-band pedestal contamination, but, within the linearized
analysis, the physical threshold does not move (the nonlinear,
strong-foam exception noted above, $\alpha\lesssim0.52$, aside).  The
progression saturates logarithmically: each decade of $\sigma(\mu)$ buys
only $\Delta\alpha=\ln10/[2\ln(D_C/\ell_P)]\approx0.008$, so the
$\sim\!3$-decade FIRAS$\to$PIXIE gain gives the $\approx0.03$ step while
PIXIE$\to$Voyage~2050 (under a decade) adds $\approx0.009$ (the forecast
entries $0.545$, $0.549$, $0.554$ round to $0.54$--$0.55$).  Even a
$10^{4}$-fold sensitivity gain over FIRAS converts to
$\Delta\alpha\lesssim0.04$, a firm ceiling: as a broadband probe based on
absolute calibrated photometry (no source-profile assumption, a different
systematic-error budget from spectroscopy), spectral-distortion
photometry yields $\alpha\gtrsim0.52$ from FIRAS and at most
$\alpha\gtrsim0.55$ next-generation, the bound limited by the
foam--$T_0$ degeneracy rather than by sensitivity or bandwidth.
In short, no absolute-photometry CMB spectrometer reaches the holographic
$\alpha=2/3$, so the high-frequency GRB channel is essential
to test holographic foam.

% ==============================================================
\section{Conclusions}
\label{sec:conclusions}

Spacetime foam imprints an achromatic, stochastic arrival-time jitter
that modifies the first-order temporal coherence $g^{(1)}(\tau)$, and
hence the observed spectrum, providing a spectroscopic probe
complementary to the imaging (PSF, Strehl) channel.  In the
dynamical-foam limit the spectrum splits into a coherent line core
suppressed by $e^{-\sigma_\phi^2}$ and an incoherent pedestal, yielding
a shape-distortion bound on the foam exponent $\alpha$.  The MeV
emission lines of GRB\,221009A give $\alpha\gtrsim 0.63$ (one-sided
$95\%$ CL), while a fit to the COBE/FIRAS monopole gives
$\alpha\gtrsim 0.52$ ($95\%$ CL); next-generation
spectrometers (PIXIE, SuperPIXIE, Voyage~2050) reach
$\alpha\gtrsim 0.54$--$0.55$, the bound saturating logarithmically
because the foam deficit is nearly degenerate with a temperature shift.
All three are dominated by this kinematic lever arm rather than by the
data: improved photon counts, line-width precision, or FIRAS-class
sensitivity would shift $\alpha$ by only $\sim0.01$--$0.02$.
The pedestal-independent line-width and CMB bounds do not reach the
holographic value $\alpha=2/3$; only the conditional TeV survival bound
($\alpha_{\rm TeV}\gtrsim0.74$, for near-Planck-refresh foam,
$\tau_{\rm foam}\lesssim4\times10^{-28}$\,s; Assumption~(ii)) would exclude it.

These bounds should be set against the window
$0.650\leq\alpha\leq0.735$ that Refs.~\cite{Steinbring2023,SteinbringNg2026}
obtain from imaging this same burst, a two-sided bracket rather than a
one-sided limit.  The two channels rest on different assumptions --- a
spectral rather than an angular observable, temporal rather than transverse
foam correlations, and the difference in reference truth noted in
Section~\ref{sec:connection} --- and relating them quantitatively requires a
common foam-cell model, so the comparison is a consistency check at the
precision those shared details allow rather than a quantitative match.  The
question it answers is whether a second route to $\alpha$ agrees with the
first, and it does.  The two pedestal-independent bounds lie below the window,
so they are floors consistent with it and with the holographic value, and they
exclude no part of it.  The conditional TeV survival bound is the single
exception, and it is the one result coupled to the imaging channel.  It
excludes the window only for near-Planck-refresh foam: if $\alpha$ lies in the
imaging window, the survival of the LHAASO photons instead requires
$\tau_{\rm foam}\gtrsim4\times10^{-28}$\,s, the pedestal having narrowed into
the band so that the survival probe closes.  An imaging detection at
$\alpha\approx2/3$ and the TeV survival of GRB\,221009A are then jointly a
statement about how fast the foam refreshes, which neither channel yields
alone.  Because the lever arm is kinematic, this agreement is a structural
feature of the two channels rather than a coincidence of present precision,
and sharpening it will require a different kind of measurement.
Because the spectroscopic and imaging channels probe the same exponent through
different observables and systematics, they form an
independent cross-check rather than a joint bracket.  The two
spectroscopic probes are themselves separated
by more than ten orders of magnitude in frequency ($\sim10^{11}$\,Hz for
the CMB versus $10^{22}$--$10^{27}$\,Hz for the GRB), and since the foam
phase variance grows as $\sigma_\phi^2\propto\nu^2$, this $\nu^2$ scaling
makes the high-frequency GRB channel intrinsically more sensitive, hence
the more constraining.  This
sensitivity advantage is partly offset by a firmer baseline on the CMB
side: its intrinsic spectrum is a near-perfect blackbody, whereas the GRB
bound depends on a less certain model of the source's intrinsic emission.

Three conditions limit the reach of these bounds.  First, all
single-source spectroscopic bounds derived in this paper assume the
dynamical-foam limit $\tau_{\rm foam}\ll\tau_c$
(Section~\ref{sec:closedforms}), in which the foam path-length jitter
decorrelates many times within each coherence time.  If foam
fluctuations are instead slow ($\tau_{\rm foam}\gg\tau_c$), the
accumulated phase is essentially frozen during each coherence time and
the foam acts as a deterministic, frequency-independent phase screen
--- no spectral broadening or peak reduction occurs within a single
observation.  The CMB constraint (Section~\ref{sec:CMB}) is derived in
the same dynamical limit ($C_\alpha\to 0$;
Section~\ref{sec:CMB_derivation}): in the quasi-static limit the
per-mode foam suppression $\mathcal{F}_{\rm foam}(\tau)\to 1$
(Eq.~\eqref{eq:Ffoam}), leaving the monopole unmodified.  In both
channels the quasi-static foam thus produces no shape distortion.
Second, a separate escape operates even for nominally dynamical foam: long-range temporal
correlations --- the temporal analogue of the transverse spatial
correlations of Ref.~\cite{LeeZurekChen2024} (Section~\ref{sec:Calpha_origin})
--- would hold $C_\alpha(\tau)\approx1$ across observational lags
regardless of photon energy, again nulling the shape-distortion signal.
A third condition is external to the foam model: the bounds 
are based on Eq.~\eqref{eq:sigmaphi_cosmo}, which does not account for the
diffractive averaging of foam structure transverse to the ray
(Section~\ref{sec:metric_origin}) --- as distinct from foam structure across
the collecting aperture, which is accounted for and is harmless when the
whole image is recorded (Appendix~\ref{app:extended}).
The proper procedure is
to compute $C(\mathbf{r},\tau)$ for candidate foam models, as done in
\citet{LeeZurekChen2024}.

The dynamical bound $\alpha\gtrsim 0.63$ (Eq.~\eqref{eq:alpha_grb_dyn})
is a weak, lever-arm-dominated floor; tightening it is limited primarily
by an instrumental gap.  The spectral region between the upper edge of
the Fermi-GBM/BGO response ($\lesssim40\,\mathrm{MeV}$) and the onset of
efficient Fermi-LAT detection ($\gtrsim100\,\mathrm{MeV}$) is poorly
covered, and the most constraining foam-induced break, at
$\nu_\star\simeq37\,\mathrm{MeV}$, falls squarely within it, so neither
instrument resolves the critical transition.
Closing this gap is a recognized community priority: next-generation
concepts (AMEGO/AMEGO-X~\cite{McEnery2019,Caputo2022},
e-ASTROGAM~\cite{DeAngelis2017}, and the near-term COSI pathfinder) target
continuous sensitivity from $\sim100\,\mathrm{keV}$ to $\gtrsim1\,\mathrm{GeV}$,
resolving the GRB spectrum across the transition for the first time.
Reducing the per-epoch line-width
uncertainties to $\delta r\lesssim 0.01$ would tighten the bound to
$\alpha\gtrsim 0.64$, and a joint centroid-plus-width fit (the present
analysis uses the width alone; Eq.~\eqref{eq:Sobs_gauss}) would add a second
handle on $\alpha$.

A complementary implication runs in the opposite direction, from a
future CMB measurement back to foam.  The
flagship goal of next-generation CMB spectrometers --- PIXIE,
SuperPIXIE, Voyage~2050 --- is a first detection of the chemical
potential $\mu$, whose guaranteed $\Lambda$CDM value
$\mu\simeq2\times10^{-8}$ arises from Silk damping of primordial
perturbations.  Our Fisher analysis shows that the foam-induced
distortion is nearly degenerate with the $\mu$ template across the
accessible band (Section~\ref{sec:CMB_degeneracy}); the two cannot be
separated from the monopole spectrum alone.  A measured $\mu$ that
departs from the $\Lambda$CDM prediction would be a landmark
signal whatever its origin, and spacetime foam is one
propagation-induced alternative to early-universe energy injection.
The GRB lower bound
$\alpha\gtrsim0.63$ already limits any foam contribution to the CMB
monopole to far below the $\Lambda$CDM $\mu$, so foam cannot
contaminate the guaranteed-$\mu$ target; conversely, were an anomalous
$\mu$ measured, the same GRB bound on $\alpha$ would be the external
input needed to test whether foam could account for it.  The decisive
limitation here is
theoretical rather than instrumental: both the pedestal shape --- which
sets how completely the foam deficit separates from $\mu$, and is fixed
by the foam temporal autocorrelation $C_\alpha(\tau)$ --- and the
amplitude $A$ are calibrated phenomenologically, still awaiting a
first-principles quantum-gravity derivation.

Finally, the multi-epoch $r_{\rm obs}$ dataset constrains any propagation
effect that broadens a line as a power of $E$, not only the
$\alpha$-foam template. The natural extension is to stochastic Lorentz-invariance
violation~\cite{AmelinoCamelia2013}, where a Planck-suppressed random
group-velocity coefficient broadens the width as
$\sigma_{\rm LIV}(E)/E\propto E^n$; unlike deterministic time-of-flight
tests~\cite{Fermi2009,LHAASOTeV2023} this width signal survives light-curve
cross-correlation. Eq.~\eqref{eq:robs_model} corresponds to the foam
case ($n=1$); refitting under $r_{\rm LIV}\propto(E/E_0)^n$ would bound
each order.  The $n=2$ case is especially valuable: deterministic
time-of-flight tests lose their lever arm at the low redshift $z=0.151$
of GRB\,221009A, so the width signal is the stronger probe there.

\section*{Data availability}
The analysis software that generates the figures and numerical results of this
paper will be deposited in a public repository with a persistent identifier upon
acceptance of this article, and is available from the authors upon reasonable
request in the interim. The COBE/FIRAS monopole spectrum and low-frequency
covariance matrix used here are public data products of the COBE/FIRAS mission,
distributed by NASA LAMBDA and described in Refs.~\cite{Fixsen1996,Fixsen2009}. The
GRB~221009A emission-line measurements analyzed here are taken from
Zhang~\textit{et al.}~\cite{YQZhang2024}.

\begin{acknowledgments}
This work was supported in part by the Director, Office of Science, Office of
High Energy Physics of the U.S.\ Department of Energy under Contract No.\
DE-AC02-05CH11231.

The authors used Claude (Anthropic) and Valency Systems to assist with drafting and editing portions of this manuscript, with detailed reference searches, and in developing the Fisher-matrix and GRB multi-epoch analysis scripts. All content, equations, numerical results, and conclusions were reviewed and verified by the authors, who take full responsibility for the work.  AI assistance was used in accordance with \textit{Physical Review} editorial policy on computational tools.

We thank Robin Kaiser for helpful discussions that helped motivate this work, and Eric Linder for insightful comments.
We also thank the anonymous referee, whose comments led to
significant improvements to the article.
\end{acknowledgments}

\bibliographystyle{apsrev4-2}
\bibliography{references}

\appendix

\section{Extended Sources and Apertures}
\label{app:extended}

The relationship between the observed and source first-order
coherence, Equation~\eqref{eq:g1obs}, was derived for a point source on a
single line of sight.  This appendix establishes the conditions under which
it holds for a spatially extended incoherent source, and for collection over
a spatially extended aperture.

Consider the source as a discrete set of points, labeled by $j$, each
seen through its own sightline with its own jitter $\delta t_j(t)$,
\begin{equation}
  E_{\rm obs}(t) = \sum_j E_{{\rm obs},j}(t),
  \qquad
  E_{{\rm obs},j}(t) \equiv E_{{\rm src},j}\!\left(t-\delta t_j(t)\right),
  \label{eq:Eobs_ext}
\end{equation}
with the points mutually incoherent,
\begin{equation}
  \big\langle E^*_{{\rm src},j}(u)\,E_{{\rm src},k}(v)\big\rangle_{\rm src}
    = \delta_{jk}\,I_j\,g^{(1)}_{{\rm src},j}(v-u),
  \label{eq:incoh}
\end{equation}
and the jitters $\{\delta t_j\}$ independent of the fields
$\{E_{{\rm src},j}\}$, as for a single sightline.  Because of that
independence the two averages may be taken in either order; we take the
source average first, at fixed jitter.

Consider first the cross terms.  Conditioning on a realization of
$\{\delta t_j\}$, the arguments of
Eq.~\eqref{eq:incoh} are fixed times, and
\begin{equation}
  \big\langle E^*_{{\rm obs},j}(u)\,E_{{\rm obs},k}(v)\big\rangle_{\rm src}
    = \delta_{jk}\,I_j\,
      g^{(1)}_{{\rm src},j}\!\left(\tau-\Delta_j(\tau)\right),
  \qquad \tau\equiv v-u,
  \label{eq:crossterms}
\end{equation}
where setting $k=j$ collapses
$[v-\delta t_k(v)]-[u-\delta t_j(u)]$ to $\tau-\Delta_j(\tau)$, with
$\Delta_j(\tau)\equiv\delta t_j(v)-\delta t_j(u)$ as in
Eq.~\eqref{eq:DeltaExt} and stationary in $\tau$.  The cross terms vanish at
fixed jitter, before the foam average is taken, so no assumption is needed
about the joint statistics of $\delta t_j$ and
$\delta t_k$.

Each sightline preserves its ensemble-mean intensity, as established
below Eq.~\eqref{eq:Eobs}, so $\langle|E_{\rm obs}|^2\rangle=\sum_j I_j$.
Averaging Eq.~\eqref{eq:crossterms} over the foam and summing over $j$,
Eq.~\eqref{eq:g1def} becomes
\begin{equation}
  g^{(1)}_{\rm obs}(\tau)
    = \frac{\sum_j I_j\,\big\langle
        g^{(1)}_{{\rm src},j}\!\left(\tau-\Delta_j(\tau)\right)
      \big\rangle_{\delta t_j}}{\sum_j I_j}
    = \frac{\sum_j I_j\displaystyle\int_{-\infty}^{\infty}
        \mathcal{N}\!\left(\Delta;0,\sigma^2_{\Delta,j}(\tau)\right)
        g^{(1)}_{{\rm src},j}(\tau-\Delta)\,d\Delta}{\sum_j I_j},
  \label{eq:g1obs_ext}
\end{equation}
the second form using the Gaussianity of Eq.~\eqref{eq:kernel}
per sightline, with $\sigma^2_{\Delta,j}$ given by Eq.~\eqref{eq:diffVarExt}
in terms of that sightline's $\sigma_\ell$ and $C_\alpha$.  This is an
intensity-weighted mean of single-sightline results, and is exact.

One further hypothesis recovers the point-source form.  Suppose the
foam statistics depend on the sightline only through the propagation
distance, and that the source subtends a spread $\delta D$ with
$\delta D/D\ll1$.  Then $\sigma^2_{\Delta,j}=\sigma^2_\Delta$ and
$C_{\alpha,j}=C_\alpha(\tau;D)$ for every $j$, the kernel is common to all
terms, and the integral commutes with the weighted sum:
\begin{equation}
  g^{(1)}_{\rm obs}(\tau)
    = \big\langle g^{(1)}_{\rm src,ext}\!\left(\tau-\Delta(\tau)\right)
      \big\rangle_{\delta t},
  \qquad
  g^{(1)}_{\rm src,ext}\equiv\frac{\sum_j I_j\,g^{(1)}_{{\rm src},j}}
                                   {\sum_j I_j}.
  \label{eq:g1obs_ext_hom}
\end{equation}
This is Eq.~\eqref{eq:g1obs} itself, with the extended source's own
intrinsic coherence $g^{(1)}_{\rm src,ext}$ in place of the point source's:
an extended incoherent source obeys the point-source result.  For a
carrier-dominated line it is
$g^{(1)}_{\rm src,ext}(\tau)\,\mathcal{F}_{\rm foam}(\tau)$; what carries
over is the foam factor, not the intrinsic profile.  Departures from the
hypothesis enter only through the sightline dependence of $\sigma_\ell$, and
hence only logarithmically in $\alpha$
(Eq.~\eqref{eq:alpha_grb_dyn}).

The construction above places the detector at a point.  A real
instrument collects the field over an aperture, and for a single source point
the foam delay may differ from one aperture point to another,
$\delta t(\mathbf{x},t)$; that transverse structure is what the imaging
channel measures.  The aperture points are not mutually incoherent --- they
carry the same source field --- so the cancellation of the cross terms above
is not available, and a different argument is needed.  It is Parseval's
theorem.  With $\tilde E(\mathbf{x},\nu)$ the temporal transform of the
aperture field and $\hat E(\mathbf{q},\nu)$ its focal-plane image,
$\int|\hat E(\mathbf{q},\nu)|^2\,d\mathbf{q}
=\int|\tilde E(\mathbf{x},\nu)|^2\,d\mathbf{x}$: the spectrum integrated over
the whole focal plane is the sum of the single-sightline spectra of
Section~\ref{sec:spec}, one per aperture point --- in the ensemble average,
each $|\tilde E|^2$ being a periodogram --- with no cross terms and no
dependence on how the foam is correlated across the aperture.

Partial collection costs less than that argument suggests.  Where the
foam is transversely uncorrelated below the aperture scale, the aperture
average of $E_{\rm src}(t-\delta t(\mathbf{x},t))$ tends to
$\langle e^{i\omega_0\delta t}\rangle E_{\rm src}(t)
=e^{-\sigma_\phi^2/2}E_{\rm src}(t)$, a constant multiple of the source field,
so the image core carries the spectrum
$e^{-\sigma_\phi^2(\nu)}S_{\rm src}(\nu)$ --- the coherent core of
Section~\ref{sec:spec} itself.  Every bound in this paper lives in that
object: the deficit fitted in Section~\ref{sec:CMB}, the survival factor of
Eq.~\eqref{eq:absolute_suppression}, and the fractional-width law, which
Section~\ref{sec:grb_width} identifies as a property of the coherent core.
What a core-only instrument excludes is the pedestal, which lies at
$\sim1/\tau_{\rm foam}$ and contributes nothing in band in any case.  This is
the sense in which the spectral and angular effects act on the same flux.

Two limitations attach.  The argument disposes of transverse foam
structure on scales between $\lambda$ and the aperture; structure below
$\lambda$ lies within a single aperture point, is the regime of
Ref.~\cite{ChenWenMa2014} discussed in Section~\ref{sec:metric_origin}, and is
untouched by Parseval.  And the instruments used here are taken to record the
image --- trivially for the FIRAS monopole, by assumption for LHAASO, where
halo containment is the issue treated in
Refs.~\cite{Steinbring2007,Steinbring2015}.  The TeV bound is conservative
either way, since flux lost to an uncollected halo only deepens the apparent
suppression.

Both extensions leave Eq.~\eqref{eq:g1obs} intact under a stated
condition: statistical homogeneity across the source, and, for the full
spectrum rather than the coherent core alone, collection over the whole
image.

\section*{Mathematical Notation}
\label{app:notation}

\begin{table}[!htb]
\caption{Principal symbols used throughout the text.}
\label{tab:notation}
\begin{ruledtabular}
\begin{tabular}{ll}
$\alpha$ & foam exponent \\
$\sigma_\ell^2(D)$ & path-length-jitter variance [Eq.~\eqref{eq:NgPerlmanVar}] \\
$\sigma_\phi^2$ & phase variance, $(2\pi/\lambda_{\rm obs})^2\sigma_\ell^2$ \\
$\sigma_\Delta^2(\tau)$ & differential-jitter variance \\
$C_\alpha(\tau;D)$ & temporal foam autocorrelation \\
$\tau_{\rm foam}$ & foam decorrelation time \\
$\tau_c\equiv\tau_G$ & coherence time of the light, $1/(2\pi\sigma_\nu)$ \\
$\sigma_{\rm obs}$ & observed line width \\
$r_{\rm obs}$ & observed fractional width, $\sigma_{\rm obs}/E$ \\
$r_{\rm int}$ & intrinsic (boost-invariant) fractional width \\
$r_{\rm foam}$ & foam-induced fractional-width excess \\
$D,\,D_C$ & (comoving) source distance \\
$\mathcal{D}$ & bulk Doppler factor \\
$A_{\rm eff}$ & lumped CMB foam amplitude \\
\end{tabular}
\end{ruledtabular}
\end{table}
\end{document}